\documentclass[aps,reprint,two column,superscriptaddress]{revtex4-2}

\usepackage[utf8]{inputenc}
\usepackage[T1]{fontenc} 
\usepackage[british]{babel} 
\usepackage[sc,osf]{mathpazo}
\usepackage[scaled=0.86]{berasans} 
\usepackage[colorlinks=true, citecolor=blue, urlcolor=blue]{hyperref} 
\usepackage{graphicx} 
\usepackage{subfig}
\usepackage[babel]{microtype} 
\usepackage{amsmath,amssymb,amsthm,bm,amsfonts,mathrsfs,bbm} 

\usepackage{xspace} 
\usepackage{pgfplots}
\usepackage{xcolor,colortbl}
\def\ba{\begin{equation}}
\def\ea{\end{equation}}
\def\bea{\begin{eqnarray}}
\def\eea{\end{eqnarray}}
\def\ben{\begin{equation*}}
\def\een{\end{equation*}}
\def\bean{\begin{eqnarray*}}
\def\eean{\end{eqnarray*}}
\def\bma{\begin{mathletters}}
\def\ema{\end{mathletters}}
\def\bi{\begin{itemize}}
\def\ei{\end{itemize}}

\newcommand{\be}{\begin{equation}}
\newcommand{\ee}{\end{equation}}

\newcommand{\kommentar}[1]{}

\newcommand{\forget}[1]{}

\begin{document}

\title{Role of Steering Inequality In Quantum Key Distribution Protocol}
\author{Kaushiki Mukherjee}
\email{kaushiki.wbes@gmail.com}
\affiliation{Department of Mathematics, Government Girls' General Degree College, Ekbalpore, Kolkata-700023, India.}
\author{Tapaswini Patro}
\email{p20190037@hyderabad.bits-pilani.ac.in}
\affiliation{Department of Mathematics, Birla Institute of Technology and Science Pilani, Hyderabad Campus,Telangana-500078, India}
\author{Nirman Ganguly}
\email{nirmanganguly@hyderabad.bits-pilani.ac.in}
\affiliation{Department of Mathematics, Birla Institute of Technology and Science Pilani, Hyderabad Campus,Telangana-500078, India}

\begin{abstract}
Violation of Bell's inequality has been the mainspring for secure key generation in an entanglement assisted Quantum Key Distribution(QKD) protocol. Various contributions have relied on the violation of Bell inequalities to build an appropriate QKD protocol. Residing between Bell nonlocality and entanglement, there exists a hybrid trait of correlations, namely correlations exhibited through the violation of steering inequalities. However, such correlations have not been put to use in QKD protocols as much as their stronger counterpart, the Bell violations. In the present work, we show that the violations of the CJWR(E.G.Cavalcanti,S.J. Jones,H.M Wiseman and M.D. Reid, Phys.Rev.A 80,032112(2009))steering inequalities can act as key ingredients in an entanglement assisted QKD protocol. We work with arbitrary two qubit entangled states, characterize them in accordance with their utility in such protocols. The characterization is based on the quantum bit error rate and violation of the CJWR inequality. Furthermore, we show that subsequent applications of local filtering operations on initially entangled states exhibiting non violation, lead to violations necessary for the successful implementation of the protocol. An additional vindication of our protocol is provided by the use of absolutely Bell-CHSH local states, states which remain Bell-CHSH local even under global unitary operations.
\end{abstract}

\maketitle

	
\section{Introduction}\label{intro}
Quantum Cryptography(QC) promises to bring a paradigmatic change in the domain of secure information processing\cite{QC1}.The state of the art techniques,recently conceptualized,have led to profound implications in how we deal with secure messages\cite{QC2}.The archetypal manifestation of quantum mechanics, namely quantum entanglement lies at the root of QC. Quantum entanglement \cite{hor1} makes it possible to realize protocols unimaginable in classical information theory.
\par An inherent constitution of QC is key distribution, which in the simplest scenario can be interpreted as the task of generating a private key between two honest users who can communicate with each other over public channels. If one includes quantum resources for secure key generation, it can outperform protocols availing only classical resources. Such a difference stems from the fact that quantum protocols basically rely on inherent random nature of quantum particles contrary to dependence on pseudo randomness and computational complexity by classical protocols\cite{retsispg10}.
\par The seminal work by Bennett and Brassard\cite{bb84} single handedly pioneered the study on quantum key distribution protocols.In the protocol described in \cite{bb84},commonly known as the BB84 protocol, the authors designed a bipartite key distribution protocol based on the idea of conjugate bases \cite{wies}, with the involvement of the protagonists Alice(sender) and Bob(receiver). Measurement induced disturbance in quantum systems\cite{nie}plays a key role in the protocol, which depends on \textit{preparation and measurement scheme}. Several key generation protocols\cite{cr,cr1,cr2,cr3,cr4,cr5,cr6} based on this scheme have been designed since the BB84 protocol.
\par A departure from the \textit{preparation and measurement scheme} is provided by the \textit{entanglement assisted key distribution} protocols. The two parties now share an entangled state\cite{hor1} and the presence of the eavesdropper is detected through the violation of a suitable Bell's inequality\cite{bell}, like the Bell-CHSH \cite{cla}. Such a scheme was envisaged by Ekert through his novel contribution\cite{eck}. Since its inception,strategies based on violation of Bell's inequalities have been used to design many QKD protocols\cite{eck1,eck2,eck3,eck4,eck5,eck6,eck7}.
\par In entanglement assisted protocols, comparison of Alice and Bob's information content with that of an eavesdropper is the most obvious way to verify security\cite{eck8}. However, violation of a suitable Bell's inequality provides a pragmatic alternative \cite{eck10,eck9,eck11,eck12}. Although, Bell violation is necessary for security,it has been shown that it is not sufficient \cite{eck14,eck13}. Consequent to a Bell violation, it becomes imperative to check the utility of the entangled state in successful key generation.In \cite{main}, the authors have characterized two qubit states in this perspective, where the relation between Bell-CHSH violation and quantum bit error rate(QBER)\cite{eck8} has been exploited.
\par Nonlocal correlations being the mainstay of QKD protocols, necessitate exploitation of other manifestations of it. Quantum steering provides for one such significant alternative, which remains sandwiched between entanglement and Bell nonlocal correlations. Schrodinger, pioneered the concept of steering \cite{str1,str2}, whose operational significance came much later through the contributions in \cite{wis,jon,red1}. A series of work followed, detecting steerability of correlations\cite{red2,red3,red4,zow,red5,san,ver,jev,sjj,costa}.
\par In the study of entanglement assisted QKDs, the role of Bell nonlocality has seen multiple probes. In this perspective, the role of other nonclassical correlations beyond Bell-CHSH inequality warrants attention.
In order to detect whether any bipartite state($\rho$,say) is steerable, we need a steering criterion. In \cite{red3}, the authors derived a linear inequality, based on linear functions of expectation values of observables, commonly referred to as CJWR inequality(for more details see subsec.\ref{steering}). A closed form (based on correlation tensor of the density matrix) to establish the violation of the said inequality, was derived in \cite{costa}. We have considered three measurement settings as in the two measurement scenario it is equivalent to Bell-CHSH inequality. States which violate the CJWR inequality under three measurement settings are also commonly termed as $ F_3 $ steerable. In the present work, we address the following problem:
\textit{Given that a quantum state violates the CJWR inequality, can it be used in a QKD protocol?}.
\par Our work, which considers an entanglement assisted key distribution protocol, makes two very significant assumptions:(i)violation of CJWR inequality for three settings is necessary for security,(ii) minimum secure key rate\cite{eck8i} is dependent on quantum bit error rate only. Precisely speaking, we characterize the two qubit state space based on the utility of a $ F_3 $ steerable state in a QKD protocol.
\par The above probe raises another imperative query, namely \textit{can the useless states be made useful for QKD with an enhancement in the protocol?}. We provide an affirmative answer, through the application of local filtering operations \cite{pop1,pop2,pop3}. Analogous to the procedure which was obtained in \cite{main} with respect to the Bell-CHSH inequality, we modify our protocol to include local filtering operations to enhance the suitability of the otherwise useless states. Strikingly, it is also observed that some $F_3$ unsteerable two qubit states can also be used for secure key generation in this modified protocol.
\par Usually, in any entanglement assisted QKD protocol, state shared between Alice and Bob is generated from some unknown source which can be under the control of an eavesdropper.
If the source distributes an absolutely Bell-CHSH local entangled state\cite{verstraeteglobal,guoch}, then secure key generation becomes impossible if Bell-CHSH violation is considered.In conjunction with this, we propose a scenario where such states can also be made useful as steerability is a weaker form of nonlocality than Bell nonlocality.\\
\par Rest of our work is organized as follows. Motivation underlying the present discussion is provided in sec.\ref{mot} followed by discussions on some mathematical pre-requisites in sec.\ref{pre}. Entire characterization of two qubit states is made in sec.\ref{res}. In sec.\ref{fils}, effect of local filtering operations in our QKD protocol is discussed. In sec.\ref{abs}, we discuss the case where absolute Bell-CHSH local states are used in the protocol. We end our discussion with some concluding remarks in sec.\ref{conc}.\\
\section{Motivation}\label{mot}
Over the years, violation of Bell-CHSH inequality has been used in analyzing security in entanglement assisted QKD protocols. Recently, in \cite{main}, considering QBER as a metric of security analysis, the authors have completely characterized arbitrary two qubit states based on their utility in such protocols.
In this context, it may be noted that violation of a suitable steering inequality may be more helpful in exploiting potential of two qubit states in the protocol. To be more precise, there may exist entangled states which cannot be used in QKD protocols relying on Bell-CHSH violation but turn out to be useful in QKD protocols involving violation of steering inequality. Such an intuition stems from the existing hierarchy of nonclassical correlations. From this perspective, it will be interesting to analyze the two qubit state space based on the violation of an appropriate steering inequality. This basically motivates the present work. Using QBER as the metric of security analysis, we have provided complete characterization of arbitrary two qubit state based on its utility in an entanglement assisted protocol involving the violation of a steering inequality.We have used the CJWR inequality \cite{red3} and the closed form for its violation\cite{costa}.Based on our analysis, it can now be checked whether a given two qubit state is useful in such a protocol or not. Also, our findings will help to point out existence of states useful in our protocol which are however useless if QKD protocol involves violation of CHSH inequality.
\par At this point, it may be noted that prior to the present work, concept of steering has been studied in the light of QKD protocols\cite{bran1,temps}. In \cite{bran1} the authors established a link between security of bipartite entanglement assisted one-sided device independent QKD scenario(only one of the two parties has trusted measurement devices) and the demonstration of quantum steering. Establishment of a steering inequality from the upper bound of secret key rate is the main result of their paper. In \cite{temps}, concept of temporal steering is used for the same in preparation and measurement based QKD schemes. However, to the best of the authors' knowledge,violation of a steering inequality pertaining to a state's efficacy in QKD protocols has not been probed earlier.
In the present submission we exploit the violation of the CJWR inequality to identify useful states in the QKD protocol. Precisely, we use the closed form derived in \cite{costa} to detect such violation. Local filtering has been used to turn some useless states(in the context of the protocol) to useful ones. Besides, some entangled states(absolutely Bell-CHSH local states\cite{verstraeteglobal,guoch}) turn out to be useful while considering the notion of $F_3$ steerability instead of Bell-CHSH violation.
\section{Preliminaries and Notations}\label{pre}
In this section, we put forward the notations to be used in our analysis with a revisit of some preliminary notions crucial to our analysis. \\
\subsection{Bloch Matrix Representation}
The density matrix $\rho$ denotes an arbitrary two qubit state shared between two parties and is given by,
\begin{equation}\label{st4}
\small{\rho}=\small{\frac{1}{4}(\mathbb{I}_{2}\times\mathbb{I}_2+\vec{\mathfrak{a}}.\vec{\sigma}\otimes \mathbb{I}_2+\mathbb{I}_2\otimes \vec{\mathfrak{b}}.\vec{\sigma}+\sum_{j_1,j_2=1}^{3}w_{j_1j_2}\sigma_{j_1}\otimes\sigma_{j_2})},
\end{equation}
with $\vec{\sigma}$$=$$(\sigma_1,\sigma_2,\sigma_3), $ $\sigma_{j_k}$ denoting Pauli operators along three mutually perpendicular directions($j_k$$=$$1,2,3$). $\vec{\mathfrak{a}}$$=$$(x_1,x_2,x_3)$ and $\vec{\mathfrak{b}}$$=$$(y_1,y_2,y_3)$ denote local bloch vectors($\vec{\mathfrak{a}},\vec{\mathfrak{b}}$$\in$$\mathbb{R}^3$) corresponding to party $\mathcal{A}$ and $\mathcal{B}$ respectively with $|\vec{\mathfrak{a}}|,|\vec{\mathfrak{b}}|$$\leq$$1$ and $(w_{i,j})_{3\times3}$ stands for the correlation tensor matrix $\mathcal{W}$(real matrix).
Components $w_{j_1j_2}$ of $\mathcal{W}$ are given by $w_{j_1j_2}$$=$$\textmd{Tr}[\rho\,\sigma_{j_1}\otimes\sigma_{j_2}].$ \\
$\mathcal{W}$ can be diagonalized by applying suitable local unitary operations\cite{gam,luo},where the simplified expression is then given by:
 \begin{equation}\label{st41}
\small{\rho}^{'}=\small{\frac{1}{4}(\mathbb{I}_{2}\times\mathbb{I}_2+\vec{\mathfrak{m}}.\vec{\sigma}\otimes \mathbb{I}_2+\mathbb{I}_2\otimes \vec{\mathfrak{n}}.\vec{\sigma}+\sum_{j=1}^{3}\mathfrak{t}_{jj}\sigma_{j}\otimes\sigma_{j})},
\end{equation}
Correlation tensor in Eq.(\ref{st41}) is given by $T$$=$$\textmd{diag}(t_{11},t_{22},t_{33})$ where $t_{11},t_{22},t_{33}$ are the eigen values of $\sqrt{\mathcal{W}^{T}\mathcal{W}},$ i.e., singular values of $\mathcal{W}.$
\subsection{Entanglement Assisted Bipartite QKD Protocol}\label{qkd}
Consider any entanglement assisted quantum key distribution(QKD) protocol\cite{eck,main} involving two parties Alice($A$) and Bob($B$), who try to establish a secure key at the end of the protocol. Let a source $\Lambda$(unknown to Alice and Bob) distribute copies of a bipartite entangled state $\rho$ between the two parties. Now Alice and Bob both perform local measurements on their respective subsystems and record their outcomes. For local measurements, each of them chooses randomly from a collection of $N$ number of $d$-dimensional basis. Let $\mathfrak{C}_{A(B)}$$=$$\{\mathfrak{B}_{A(B)}^{(\beta)}\}_{\beta=1}^{N}$ denote the collection of $N$ bases of Alice(Bob) where $\forall\,\beta,$ $\mathfrak{B}_{A(B)}^{(\beta)}$ are given by:
\begin{equation}\label{basis1}
    \mathfrak{B}_{A}^{(\beta)}=\{|\psi^{\beta}_i\rangle\}_{i=1}^d\,\textmd{and}
    \mathfrak{B}_{B}^{(\beta)}=\{|\phi^{\beta}_i\rangle\}_{i=1}^d
\end{equation}
If $\mathcal{O}_{A(B)}^{(\beta)}$ denote operators corresponding to the basis $\mathfrak{B}_{A(B)}^{(\beta)}$,then those are given by:
\begin{equation}\label{opes}
    \mathcal{O}_{A(B)}^{(\beta)}=\{|\psi(\phi)^{\beta}_i\rangle\langle \psi(\phi)^{\beta}_i|\}_{i=1}^d,\,\forall \beta=1,...,N.
\end{equation}
Now the bases of Alice and Bob $\mathfrak{C}_{A}$ and $\mathfrak{C}_{B}$ are correlated in the following sense. Let for any fixed value of $\beta$ (from $1,2,...,N$), Alice and Bob performs measurement in $\mathfrak{B}_A^{(\beta)}$ and $\mathfrak{B}_B^{(\beta)}$ respectively. In case $\rho$(shared between them) is a pure entangled state, then perfect correlations(between Alice and Bob's outputs) will imply that if Alice gets outcome $|\psi^{\beta}_j\rangle$ then Bob's outcome must be $|\phi^{\beta}_j\rangle(\forall\,j).$ \\
After performing measurements on $N$ copies of $\rho,$ a fraction of the measurement outcomes are used to analyze the joint statistics for verifying whether corresponding correlations are nonlocal. Such a verification is made by testing violation of a Bell inequality. For the remaining part of the measurement outcomes, the parties publicly compare their measurement bases and keep outcomes only corresponding to the correlated bases(discarding the remaining outcomes). These outcomes(obtained from correlated bases) form the raw key\cite{bran1}. The parties can then extract secure key from remaining part of the raw key by performing information reconciliation\cite{eck8i,in1} and privacy amplification\cite{eck8i}.
\subsubsection{Quantum Bit Error Rate}\label{qber}
For any given state $\rho,$ QBER($Q$) is defined as the average mismatch between Alice and Bob's outcomes obtained when they measure in correlated bases. With $\mathfrak{C}_{A}$ and $\mathfrak{C}_{B}$ denoting collection of correlated bases(Eq.(\ref{basis1})) of the two parties(as considered above), QBER can be expressed as:
\begin{equation}\label{basis2}
    Q=\frac{1}{N}\sum_{\beta=1}^{N}\sum_{i\neq j=1}^d \langle \psi^{\beta}_i\phi^{\beta}_j|\rho|\psi^{\beta}_i\phi^{\beta}_j\rangle
\end{equation}
The above expression of $Q$ holds for any $N$$\leq$$d+1$ number of bases. For instance, when source $\Lambda$ generates a two qubit state($d$$=$$2$) and each party chooses from a collection of two bases, i.e., $|\mathfrak{C}_{A}|$$=$$|\mathfrak{C}_{B}|$$=$$2,$ QBER is given by\cite{main}:
\begin{equation}\label{basis3}
    Q=\frac{1}{4}(2-\vec{u_1}.\mathcal{W}\vec{v_1}-\vec{u_2}.\mathcal{W}\vec{v_2})
\end{equation}
where $\vec{u}_i,\vec{v}_j\,(i,j$$=$$1,2)$ denote Bloch vectors of the measurement bases of Alice and Bob respectively and $\mathcal{W}$ denotes the correlation tensor(Eq.(\ref{st4})). Minimization over all possible measurement directions $\vec{u}_1,\vec{u}_2,\vec{v}_1,\vec{v}_2$ gives:
\begin{equation}\label{basis4}
    Q\geq \frac{1}{4}(2-\textmd{max}_{i,j}(|t_{ii}|+|t_{jj}|)),\,i\neq j
\end{equation}
where $t_{11},t_{22},t_{33}$ denote the singular values of correlation tensor $T$ of $\rho^{'}$(Eq.(\ref{st41})) and hence singular values of correlation tensor $\mathcal{W}$ of $\rho$(Eq.(\ref{st4})).

\subsection{Steering Inequality}\label{steering}
 Linear steering inequalities, based on linear functions of expectation values of observables, provide useful way to detect the steerability of a state. In general, if a given state in finite dimensions is steerable, then exists a linear criterion to exhibit steering \cite{red3}.
 In \cite{red3}, a linear steering inequality was formulated,under the assumption that both the parties(Alice and Bob) sharing a bipartite state($\rho$) perform $n$ dichotomic quantum
measurements(on their respective particles). Cavalcanti, Jones, Wiseman, and Reid(CJWR) derived a
series of correlators based inequalities\cite{red3} for verifying steerability of $\rho:$
\begin{equation}\label{mon2}
\mathcal{F}_{n}(\rho,\nu) = \frac{1}{\sqrt{n}}|\sum_{l=1}^{n} \langle A_{l} \otimes B_{l} \rangle | \leq 1
\end{equation}
where\\
$A_{l} $$=$$ \hat{a}_{l}$$\cdot$$ \overrightarrow{\sigma},$ $B_{l}$$=$$\hat{b}_{l}$$\cdot$$ \overrightarrow{\sigma}$ with\\
$\hat{a}_{l} $$\in$$ \mathbb{R}^{3}\, \textmd{being unit vectors whereas}\,\hat{b}_{l} $$\in$$ \mathbb{R}^{3}\,$denote orthonormal vectors$.\,\nu $$=$$\{\hat{a}_{1},\hat{a}_{2},....\hat{a}_{n}, \hat{b}_{1},\hat{b}_{2},...,\hat{b}_{n} \}$ stands for the collection of measurement directions, $\langle A_{l} \otimes B_{l} \rangle$$=$$ \textmd{Tr}(\rho (A_{l} \otimes B_{l}))$ and $\rho$$\in$$ \mathbb{H_{A}} $$\otimes$$ \mathbb{H_{B}}$ is any bipartite quantum state. Violation of Eq.(\ref{mon2}) ensures both way steerability of $\rho$ in the sense that it is steerable from $A$ to $B$ and vice versa. In particular, for $n$$=$$3,$ CJWR inequality(Eq.(\ref{mon2})) for three settings takes the form:
\begin{equation}\label{mon2i}
\mathcal{F}_{3}(\rho,\nu) = \frac{1}{\sqrt{3}}|\sum_{l=1}^{3} \langle A_{l} \otimes B_{l} \rangle | \leq 1
\end{equation}
\\
In \cite{costa}, analytical expressions for the upper bound of CJWR steering
inequality were formulated in terms of correlation tensor parameters of $\rho.$ Analytical expression of the upper bound of corresponding inequality(Eq.(\ref{mon2i})) is given by:
\begin{eqnarray}\label{string}
\textmd{Max}_{\nu}\mathcal{ F}_{3}(\rho,\nu)=\sqrt{\textmd{Tr}[\mathcal{W}^T\mathcal{W}]}
\end{eqnarray}
where $\mathcal{W}$ denote the correlation tensor corresponding to bloch matrix representation of $\rho$(Eq.(\ref{st4})). So, by the linear inequality(Eq.\ref{mon2i}), any two-qubit state $\rho$(Eq.(\ref{st4})), shared between $A$ and $B$ is both-way $F_{3}$ steerable if:
\begin{equation}\label{mon8}
\sqrt{Tr[\mathcal{W}^{T} \mathcal{W}]}>1.
\end{equation}
\subsection{Local Filtering Operations}\label{locf}
Local filtering operations form a special class of sequential quantum operations\cite{pop1,pop2,pop3}. By applying suitable local filtering operations, the entanglement concentration and nonlocal content of any state $\rho$ can be increased. Let $M_{A(B)}^{(1)},$ $M_{A(B)}^{(2)}$ denote the local filtering operations applied by Alice(Bob) on their respective subsystems. Under application of these filtering operation state $\rho$ gets transformed to a new state $\rho^{'}$\cite{pop2,pop3}:
\begin{equation}\label{filter1}
    \rho^{'}=\frac{M_{A}^{(1)}\otimes M_{B}^{(1)}\rho (M_{A}^{(1)}\otimes M_{B}^{(1)})^{\dagger}}
    {\textmd{Tr}[M_{A}^{(1)}\otimes M_{B}^{(1)}\rho( M_{A}^{(1)}\otimes M_{B}^{(1)})^{\dagger}]}
\end{equation}
For simplicity, we consider $M_{A(B)}^{(2)}$$=$$\sqrt{\mathbb{I}_2-M_{A(B)}^{(1)}}$ with the following specific forms of $M_{A(B)}^{(1)}:$
\begin{eqnarray}\label{filter2}
  M_{A}^{(1)} &=& \epsilon_1|0\rangle\langle0|+|1\rangle\langle 1|\\
  M_{B}^{(1)} &=& \epsilon_2|0\rangle\langle0|+|1\rangle\langle 1|,\,\textmd{with}\,\epsilon_1,\epsilon_2\in [0,1]
\end{eqnarray}
It may be noted here that local filtering operations for two qubit states may be considered as single copy entanglement distillation operations\cite{pop2,pop3}.
\subsection{Absolutely Bell-CHSH local states}
An intriguing status is presented by some quantum states which remain Bell-CHSH local even under the application of global unitary operations \cite{verstraeteglobal,guoch}. Such states are termed as absolutely Bell-CHSH local \cite{guoch}.
\par If $ a_1,a_2,a_3,a_4 $ are eigenvalues of a two qubit density matrix in descending order $ a_1 \ge a_2 \ge a_3 \ge a_4 $, then the state is absolutely Bell-CHSH local iff,
\begin{equation}
	(2a_1+2a_2-1)^2 + (2a_1+2a_3-1)^2 \le 1
\end{equation}
\section{Characterizing arbitrary two qubit states based on $Q$}\label{res}
In this section, we characterize any given quantum state with respect to its utility in a QKD protocol. Before starting our analysis, we first discuss the scenario in detail.\\
\subsection{Measurement Specifications}\label{pros}
For our purpose we consider the usual bipartite entanglement assisted QKD protocol(subsec.\ref{qkd}) such that $\rho$ shared between Alice and Bob is a two- qubit state($d$$=$$2$). At this junction, one may note that excepting the dimensionality($2$ in this case), none of Alice and Bob has any other information about $\rho.$ In this protocol, each of the parties performs local measurements on their respective subsystems. Alice chooses randomly from a collection of three projective measurements in arbitrary directions $\mathfrak{C}_{A}$$=$$\{\mathfrak{B}_{A}^{(\beta)}\}_{\beta=1}^{3}.$ Bob chooses randomly from a collection of three mutually unbiased bases(MUBs\cite{mub1}): $\mathfrak{C}_{B}$$=$$\{\mathfrak{B}_{B}^{(\beta)}\}_{\beta=1}^{3}.$ $\mathfrak{B}_{A(B)}^{(\beta)}$ are given by Eq.(\ref{basis1}) for $d$$=$$2.$ Bases of Bob being mutually unbiased\cite{mub1}, $\langle \phi_i^{\beta}|\phi_{i^{'}}^{\beta^{'}}\rangle$$=$$\frac{1}{\sqrt{2}}$ where $i,i^{'}$$\in$$\{0,1\}$ and $\beta$$\neq$$\beta^{'}.$ Alice cannot use MUBs as this will make her measurements characterized. After making measurements, the parties reconcile their measurement bases publicly. Now with $x,y$ denoting measurements and $a,b$ denoting outcomes of Alice and Bob respectively, the correlation statistics $P(a,b|x,y),$ corresponding to a fraction of raw data(measurement outcomes) is used for checking $F_3$ steerability of $\rho$ via the violation of CJWR inequality for three settings(Eq.\ref{mon2i}). Both Alice and Bob perform projective measurements in arbitrary directions: $\vec{u}_i.\vec{\sigma}$ and $\vec{v}_i.\vec{\sigma}\,(i$$=$$1,2,3)$ respectively. $\forall i,\,\vec{u}_i$ and $\vec{v}_i$ represent bloch vectors of measurement basis with Alice and Bob respectively. Each party choosing from a collection of three bases implies that each of them can choose to perform projective measurement in any one of three arbitrary directions: $\vec{u}_1,\vec{u}_2,\vec{u}_3$ for Alice and $\vec{v}_1,\vec{v}_2,\vec{v}_3$ for Bob. While $\vec{u}_1,\vec{u}_2,\vec{u}_3$ are only unit vectors, $\vec{v}_1,\vec{v}_2,\vec{v}_3$ are orthonormal vectors so that Bob's measurements are mutually unbiased qubit measurements. Having specified the measurement scenario, we next approach to optimize the QBER($Q$) for our scenario. \\
\subsection{Optimization of $Q$}\label{opt}
In a QKD protocol involving three measurement settings per party, $Q$(Eq.(\ref{basis2})) turns out to be:
\begin{equation}\label{basis6}
     Q=\frac{1}{6}(3-\sum_{i=1}^3\vec{u_i}.\mathcal{W}\vec{v_i})
\end{equation}
where $\mathcal{W}$ is the correlation tensor appearing in Bloch matrix representation of $\rho$(Eq.(\ref{st4})).
Minimization over all possible measurement directions $\vec{u}_1,\vec{u}_2,\vec{u}_3,\vec{v}_1,\vec{v}_2,\vec{v}_3$ gives(see appendix Sec.\ref{app1}):
\begin{eqnarray}\label{basis7}
Q&\geq &Q_{min},\, \textmd{where},\,\nonumber\\
Q_{min}&=&\frac{1}{6}(3-\sum_{i=1}^3t_{ii}).
\end{eqnarray}
where $T$$=$$\textmd{diag}(t_{11},t_{22},t_{33})$ denote correlation tensor of $\rho^{'}$(Eq.(\ref{st41})). As discussed in subsec.\ref{pre}, $t_{11},t_{22},t_{33}$ are the singular values of correlation tensor $\mathcal{W}$ of $\rho$(Eq.(\ref{st4})).
Now let $\rho$ be an $F_3$ unsteerable state. For simplicity, let singular values of its correlation tensor($\mathcal{W}$) satisfy:
\begin{equation}\label{basis7i}
   t_{11}^{2}+ t_{22}^{2}+t_{33}^2=1
\end{equation}
Now we minimize $Q$ with respect to all such possible $F_3$ unsteerable quantum states. Imposition of such a restriction is required as we are considering violation of CJWR inequality(Eq.(\ref{mon2i})) by $\rho$ necessary for successful key generation in the protocol(using $\rho$). If $\mathcal{Q}_0$ denote the least possible value of $Q$ under such restriction(see appendix \ref{app2}), then:
\begin{equation}\label{bases7}
    \mathcal{Q}_0=0.211.
\end{equation}
Consequently, when any $F_3$ unsteerable state is used for key distribution, QBER generated in the protocol cannot be less than $0.211.$ The minimum error rate($\mathcal{Q}_0$) may also be referred to as the
\textit{critical error rate} of our QKD protocol. However when any $F_3$ steerable state is used, QBER can be less than $\mathcal{Q}_0$(to be discussed in subsec.\ref{char}). Critical error rate($\mathcal{Q}_0$) is obtained for $t_{11}$$=$$t_{22}$$=$$t_{33}$$=$$\frac{1}{\sqrt{3}}$(see appendix \ref{app2}). \\
Having obtained the critical value of QBER($\mathcal{Q}_0$), we precisely list down the steps used to check security of our protocol\\
\subsection{Steps of the QKD Protocol}
Consider that $N$ copies of a two qubit state $\rho$ are generated from a source and distributed between Alice and Bob. Each of them thus receives $N$ qubits(one from each copy of $\rho$). \\
\textit{Step.1:} Some of $N$ copies, for instance, say $k_1$$<$$N$ are used to test CJWR inequality. For that the parties perform local projective measurements on their respective qubits( as discussed in subsec.\ref{pros}). If corresponding statistics do not violate the inequality then the protocol is aborted. If violation is observed then the users perform the next step.\\
\textit{Step.2:} They measure remaining $N-k_1$ copies in local bases. Using classical communication, they compare their bases and keep the outputs corresponding to correlated bases only. A portion of those measurement statistics is then used to calculate QBER. If QBER exceeds $\mathcal{Q}_0,$ protocol is aborted. Else the remaining outputs corresponding to the correlated bases are used for secure key generation.\\
 We next characterize two qubit state space in context of secure key generation.\\
\subsection{Characterization of Two Qubit State Space}\label{char}
\begin{center}
\begin{figure}
\includegraphics[width=3.3in]{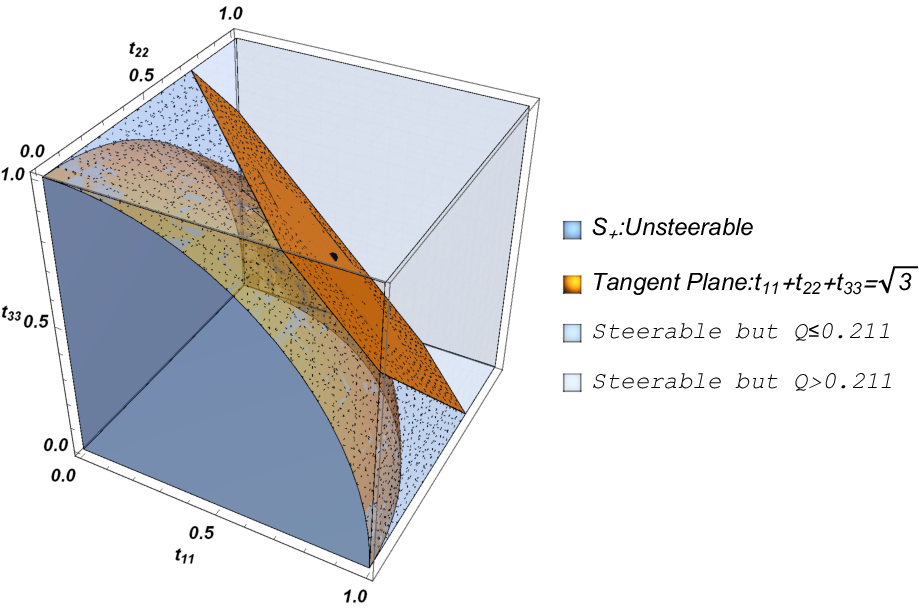} \\
 \caption{\emph{ Singular value space of correlation tensor of an arbitrary two- qubit state considered here. Cuboid indicates all possible two-qubit states whereas any point from the first octant $S_{+}$ of sphere $S$(Eq.(\ref{basis7i}))gives $F_3$ unsteerable state. Region lying outside $S_+$ and below the tangent plane at $P(\frac{1}{\sqrt{3}},\frac{1}{\sqrt{3}},\frac{1}{\sqrt{3}})$ indicates steerable but useless states whereas that lying above the tangent plane gives useful states.}}
\label{fig}
 \end{figure}
 \end{center}

As already discussed before, here we intend to characterize an arbitrary two qubit state $\rho$ pertaining to its utility in the QKD scenario discussed in subsec.\ref{qkd}. As is evident from our discussion so far in sec.\ref{res}, analyzing the singular value space of correlation tensor $\mathcal{W}$(Eq.(\ref{st4}))
suffices for our purpose. \\
$\rho$ being a quantum state, in general, each of $t_{11},t_{22},t_{33}$$\in$$[0,1].$ Let $C$ denote a unit cuboid:
\begin{equation}\label{basis8i}
    C=\{(t_{11},t_{22},t_{33}):0\leq t_{11},t_{22},t_{33}\leq1\}
\end{equation}
So density matrix corresponding to any point lying outside the cuboid $ C $(Eq.(\ref{basis8i})) does not represent any valid quantum state(see Fig.\ref{fig}). For rest of our analysis, we denote quantum state corresponding to any point $R$ inside $C$ as $\rho_{R}.$ Now, consider the unit sphere($S$,say) with center at the origin given by Eq.(\ref{basis7i}). Only first octant($S_{+}$,say) of $S$ lies inside $C.$ $F_3$ unsteerable states reside on and inside $S_{+}.$ So any point lying inside $C$ but outside $S_{+}$ corresponds to an $F_3$ steerable quantum state(see Fig.\ref{fig}). Now, as discussed in subsec.\ref{opt},when an $F_3$ unsteerable state $\rho$ is used then $Q$$\geq$$\mathcal{Q}_0.$ This in turn restricts the singular values of $\mathcal{W}$(corresponding to $\rho$):
\begin{equation}\label{basis9ii}
 t_{11}+t_{22}+t_{33}\leq\sqrt{3}.
\end{equation}
Eq.(\ref{basis9ii}) represents region lying below a tangent plane to $S_{+}$ at the point $P(\frac{1}{\sqrt{3}},\frac{1}{\sqrt{3}},\frac{1}{\sqrt{3}}).$
Clearly when $\rho$ corresponding to any point lying below or on the tangent plane(Eq.(\ref{basis9ii})) is used in the protocol, estimated QBER is greater than or at most equal to the critical error rate($\mathcal{Q}_0$). On the contrary, when $\rho$ used in the protocol corresponds to any point($R$,say) lying above the same plane, QBER is less than $\mathcal{Q}_0.$ Clearly, in such a case, the point $R$ lies outside $S_{+}$(see Fig.\ref{fig}). Consequently, $\rho_R$ is $F_3$ steerable. Under our assumption of CJWR inequality's violation necessary for ensuring secure key generation in the protocol, $\rho_R$ thus turns out to be useful. In this context, we consider \textit{any state $\rho$ as useful in our QKD protocol if QBER obtained in the protocol(using $\rho$) is less than $\mathcal{Q}_0.$} Hence two qubit state $\rho$ corresponding to any point lying in $C$ is useful if and only if:
\begin{equation}\label{basis9iii}
 t_{11}+t_{22}+t_{33}>\sqrt{3}.
\end{equation}\\

\begin{figure}[htb]
\includegraphics[width=3.3in]{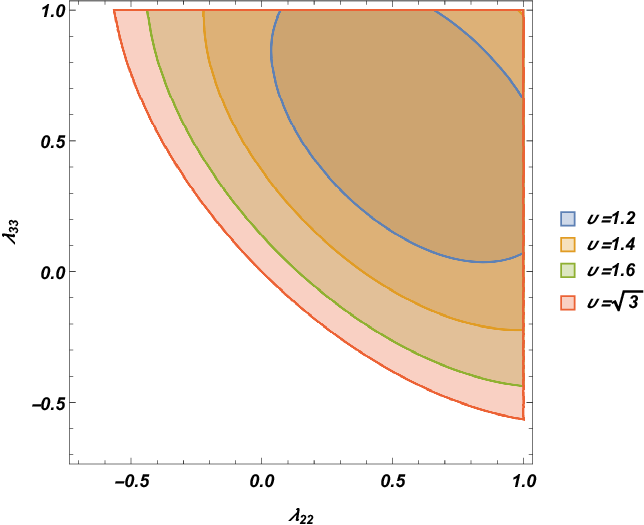}\\
\caption{\emph{Shaded regions give two qubit states useful for QKD for some specific violation amounts of CJWR inequality observed in the protocol.}}
\label{fig2}
\end{figure}

Now,let us focus on the region lying outside $S_{+}$ and inside $C.$ Let $L$ be any point lying in that region. So $\rho_L$ is $F_3$ steerable. $L$ may lie below or above the tangent plane(Eq.(\ref{basis9ii})). Utility of $\rho_L$ thus depends on position of $L.$ To be precise, if $L$ lies above the tangent plane then $\rho_L$ is useful in our protocol whereas $\rho_L$ turns out to be useless in the other case($L$ lying below the plane). This in turn points out the insufficiency of the $F_3$ steerability criterion to ensure secure key generation. Three settings CJWR inequality(Eq.(\ref{mon2i})) being a Bell type inequality, our observation simply points out the following:\\
\textit{Violation of a Bell type inequality by any two-qubit state $\rho$ is necessary but not sufficient to guarantee secure key generation in an entanglement assisted protocol involving $\rho.$ }\\
In practical situations owing to unavailability of pure entangled state for key distribution, observation of maximal violation of CJWR inequality becomes impossible. Hence, based on the amount of violation, identifying two qubit entangled states useful in entanglement assisted QKD protocol is important from practical view point. \\
\textit{Identifying Useful States:} Let in a QKD protocol, CJWR inequality(Eq.(\ref{mon2i})) be violated by some fixed amount $\mathcal{V}$(say). Let $\varrho$ be some unknown two qubit state used in the corresponding protocol. Let $\lambda_{11},\lambda_{22},\lambda_{33}$ denote the singular values of the correlation tensor of $\varrho.$ Violation being observed, restriction is imposed on these three unknown(as $\varrho$ is unknown) quantities,namely:
\begin{eqnarray}\label{basis10}
    \lambda_{11}^2+\lambda_{22}^2+\lambda_{33}^2&=&\mathcal{V}^2\nonumber\\
    \textmd{Alternatively,}\, \lambda_{11}&=&\sqrt{\mathcal{V}^2-\lambda_{22}^2-\lambda_{33}^2}
\end{eqnarray}
Now $\varrho$ is useful for secure key generation in the protocol if it satisfies Eq.(\ref{basis9iii}). Hence, $\varrho$ is useful if:
\begin{equation}\label{basis11}
\sqrt{3}-\lambda_{33}-\lambda_{22}<\sqrt{\mathcal{V}^2-\lambda_{22}^2-\lambda_{33}^2}\leq 1
\end{equation}
Eq.(\ref{basis11}) thus specifies the criterion required to be satisfied by an unknown state provided $\mathcal{V}$ amount of violation of CJWR inequality is observed in the protocol(see Fig.\ref{fig2}).\\

\subsection{Illustrations}
\begin{figure}[htb]
\includegraphics[width=3.3in]{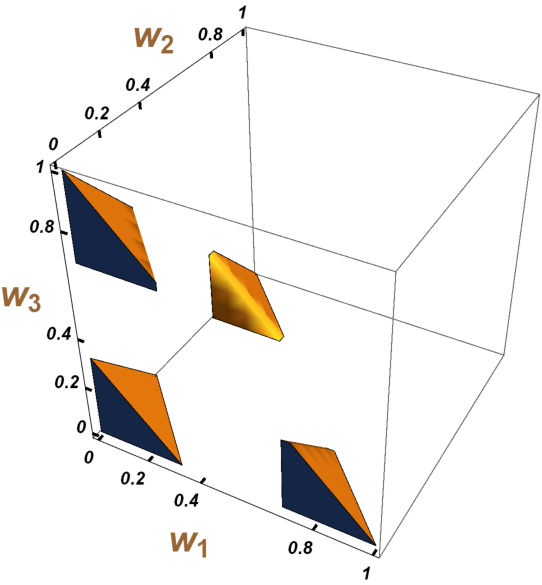}\\
\caption{\emph{Shaded region forms a part of parameter space of the Bell diagonal family(Eq.(\ref{belldiag})). Bell diagonal state corresponding to any point in the shaded region is useful in QKD protocol.}}
\label{fig3}
\end{figure}

Let us now analyze the above characterization for a few well known class of two qubit states. \\
\textit{Bell Diagonal states:} The class of Bell diagonal states\cite{nie} is represented as follows:
\begin{eqnarray}\label{belldiag}
    \varrho_{Bell}=w_1 |\psi^{-}\rangle\langle \psi^{-}| +w_2|\phi^{+}\rangle\langle \phi^{+}|+w_3 |\phi^{-}\rangle\langle \phi^{-}|\nonumber\\
    +w_4|\psi^{+}\rangle\langle \psi^{+}|&&,
\end{eqnarray}
with $w_i$$\small{\in}$$[0,1]\,\forall i$$=$$1,2,3,4,$ $\sum_{i=1}^4w_i$$=$$1$ and $ |\phi^{\pm}\rangle$$=$$\frac{|00\rangle\pm|11\rangle}{\sqrt{2}},$ $ |\psi^{\pm}\rangle$$=$$\frac{|01\rangle\pm|10\rangle}{\sqrt{2}}$ denote the Bell states. Eq.(\ref{belldiag}) is often referred to as the class of states having maximally mixed marginals.\\
Correlation matrix is given by: $\textmd{diag}(1 - 2 (w_1 + w_3),1 - 2 (w_2 + w_3),1 - 2 (w_1 + w_2))$. Bell diagonal states are $F_3$ steerable provided the following relation holds:
\begin{equation}\label{belldiag1}
   \sqrt{8(\sum_{i,j=1}^3w_i*w_j+w_4)-5}>1
\end{equation}
On the other hand, Bell diagonal states useful for QKD protocol(satisfying Eq.(\ref{basis9iii})) are characterized by:
\begin{equation}\label{belldiag2}
  \sum_{i,j=1}^3 |1 - 2 (w_i + w_j)|>\sqrt{3},\,\textmd{where}\, i\neq j
\end{equation}
Combination of Eqs.(\ref{belldiag1},\ref{belldiag2}) points out that not all $F_3$ steerable states from this family are useful for our QKD protocol(see Fig.\ref{fig3}). \\

\begin{figure}[htb]
\includegraphics[width=3.3in]{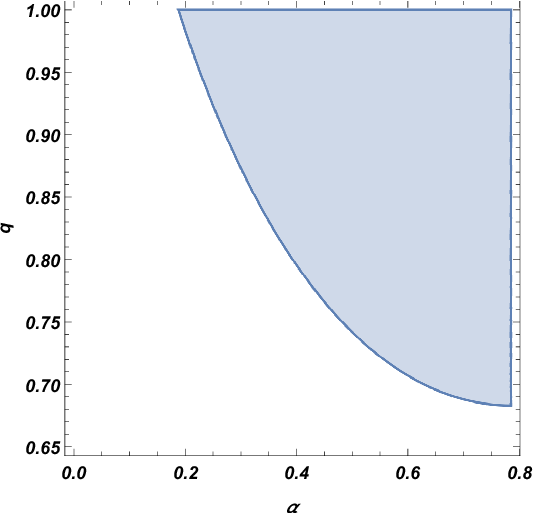}\\
\caption{\emph{Shaded region gives the states(Eq.(\ref{s14})) which can be used for secure key generation.}}
\label{fig4}
\end{figure}

Now let us consider the Werner class of states from the family of Bell diagonal states:
\begin{equation}\label{werner}
    \varrho_{W}=\omega |\psi^{-}\rangle\langle \psi^{-}|+\frac{(1-\omega)}{4}\mathbb{I}_{2}\times\mathbb{I}_2,\,\omega\in[0,1].
\end{equation}
For $\omega$$\small{\in}$$(0.5772,1]$ corresponding member from Werner class(Eq.(\ref{werner})) is $F_3$ steerable. Again Eq.(\ref{basis9iii}) is satisfied for the same range of values of $\omega$.Consequently for this subclass of Bell diagonal states(Eq.(\ref{belldiag})), any $F_3$ steerable state is always useful in QKD protocol. \\
\textit{Family of States Not Diagonal in Bell Basis:} Consider the following class\cite{grud,woj}:
\begin{equation}\label{s14}
    \gamma=q |\varphi\rangle\langle\varphi|+(1-q)|00\rangle\langle00|
\end{equation}
where $|\varphi\rangle=\cos\alpha|10\rangle+\sin\alpha|01\rangle,$ with $\alpha$$\in$$[0,\frac{\pi}{4}]$ and $0$$\leq $$q$$\leq$$1$. This class of states was used in \cite{grud} for increasing maximally entangled fraction in an entanglement swapping network. Correlation tensor is given by $\textmd{diag}(q\sin2\alpha,q\sin2\alpha,1-2q).$ A member from this family is $F_3$ steerable if:
\begin{equation}\label{s141}
    2 q^2 \sin^2 2\alpha + (1 - 2 q)^2> 1
 \end{equation}
Again any state from this class is useful in QKD protocol in case it satisfies the following relation:
\begin{equation}\label{s142}
  2q\sin2\alpha+|1-2q|>\sqrt{3}.
\end{equation}
Clearly, QKD protocol will run successfully if states satisfying both the above relations(Eqs.(\ref{s141},\ref{s142})) are used in the protocol(see Fig.\ref{fig4}).\\

\subsection{Higher Tolerance to QBER}
\begin{figure}[htb]
\includegraphics[width=3.3in]{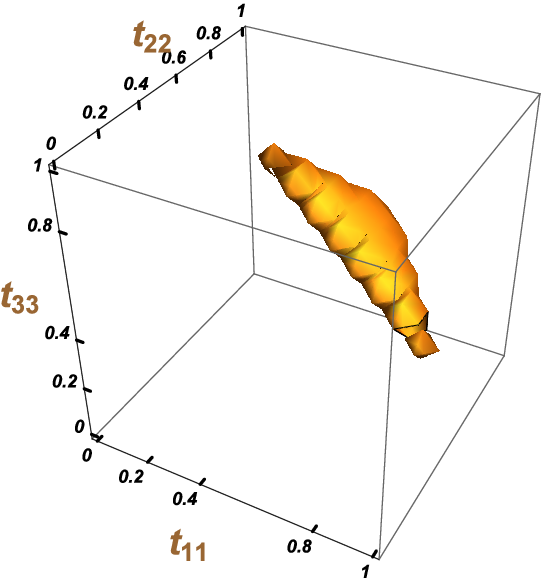}\\
\caption{\emph{Shaded region forms a part of singular value space of correlation tensor of an arbitrary two-qubit state. State corresponding to any point from this region is useless in CHSH based QKD protocol whereas the same is useful in our protocol.}}
\label{new1}
\end{figure}

\begin{figure}[htb]
\includegraphics[width=3.3in]{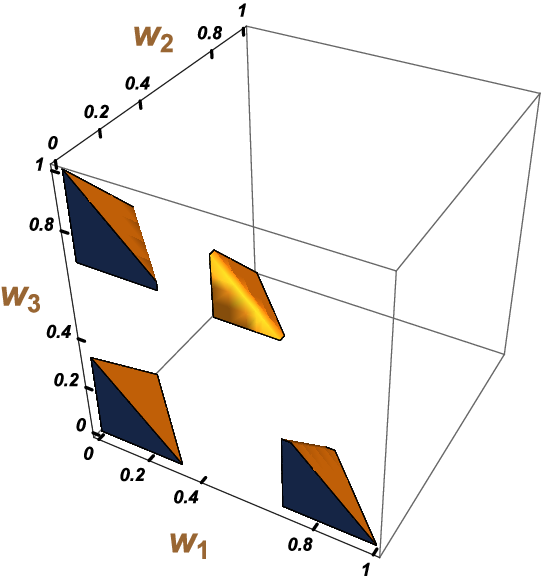}\\
\caption{\emph{Shaded part of parameter space of Bell diagonal family(Eq.(\ref{belldiag})) gives states useful in CJWR based QKD protocol but useless in CHSH based protocol.}}
\label{new2}
\end{figure}

Owing to communication of quantum states over noisy channels, non zero QBER is generated in any entanglement assisted QKD protocol even in absence of any eavesdropper. For analyzing QBER tolerance in CHSH based protocol with that in CJWR based one, we assume that the QKD protocol involves only the legitimate users(Alice and Bob), i.e., absence of any third party(eavesdropper). In \cite{main}, it was shown that an arbitrary two qubit state is useful in standard entanglement assisted QKD protocol(involving Bell-CHSH violation) if:
\begin{equation}\label{9chsh}
 \textmd{Max}\{t_{11}+t_{22},t_{33}+t_{22},t_{11}+t_{33}\}>\sqrt{2}.
\end{equation}
However the same state is useful in our protocol if it satisfies Eq.(\ref{basis9iii}). Comparison of Eqs.(\ref{9chsh},\ref{basis9iii}) points out existence of two qubit states(see Fig.\ref{new1}) satisfying Eqs.(\ref{basis9iii}) but violating Eq.(\ref{9chsh}). Let us now explore with few specific instances in this regard. \\
Consider the family of Bell diagonal states(Eq.(\ref{belldiag})). Any state from this family is not useful in the protocol relying on Bell-CHSH violation if:
\begin{equation}\label{belldiag24}
  \textmd{Max}_{i,j=1}^3 |1 - 2 (w_i + w_j)|\leq\sqrt{2},\,\textmd{where}\, i\neq j
\end{equation}
However the same state is useful in our protocol if Eq.(\ref{belldiag2}) is satisfied. There exist states from this family(see Fig.\ref{new2}) which satisfy both Eqs.(\ref{belldiag2},\ref{belldiag24}).\\
For a more specific instance from this family(Eq.(\ref{belldiag})), let us consider the Werner class of states(Eq.(\ref{werner})). Any member from this subclass of Bell diagonal states, characterized by $\omega$$\in(0.5,0.707],$ is useful in QKD protocol only if the protocol relies on violation of CJWR inequality.\\
For the family of states given by Eq.(\ref{s14}), a state is not useful in CHSH based QKD protocol if :
\begin{equation}\label{s151}
  \textmd{Max}\{2q\sin2\alpha,q\sin2\alpha+|1-2q|\}>\sqrt{2}.
\end{equation}
Comparing Eq.(\ref{s151}) with Eq.(\ref{s142}), we get states that can be used in our protocol but are useless in CHSH based ones.\\
\par As discussed in subsec.\ref{opt}, whenever an unsteerable state is used in our protocol QBER can never be less than $\mathcal{Q}_0$$=$$0.211.$ Critical value of QBER for our protocol(based on CJWR inequality) is greater than that obtained in the protocol when it relies upon Bell-CHSH inequality where $\mathcal{Q}_0$$=$$0.14$\cite{main}. Hence, for any state $\rho,$ if QBER in QKD protocol(assuming absence of any Eavesdropper) lies in range $(0.14,0.211],$ then the protocol can be used if it relies upon CJWR inequality's violation but cannot be used if it is based on Bell-CHSH violation. Consequently, the protocol turns out be more QBER tolerant when based upon the notion of steerability compared to that obtained in Bell-CHSH based QKD protocol.
\begin{figure}[htb]
\includegraphics[width=3.3in]{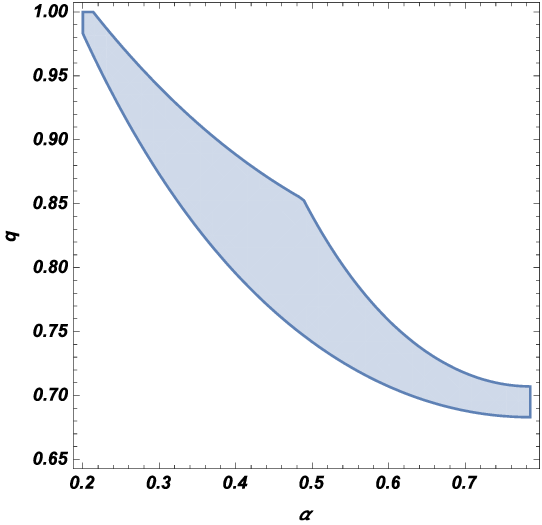}\\
\caption{\emph{Shaded part of parameter space of family of states given by Eq.(\ref{s14})) gives states useful in CJWR based QKD protocol but useless in CHSH based protocol.}}
\label{new3}
\end{figure}

\section{Incorporating Local Filtering Operations in QKD}\label{fils}
As noted before local filtering operations are crucial to enhance utility in information processing tasks \cite{such1,such2,gupta}.Let us now consider an entanglement assisted protocol where both the parties can perform local filtering operations before measuring their subsystems in correlated bases. We next discuss the protocol in detail.\
\subsection{Modified QKD Protocol}
Steps of the protocol are specified below:\\
\begin{itemize}
\item Source $\Lambda$ distributes many copies of a two qubit state($\rho$) between the two parties Alice and Bob. \\
\item On receiving the qubits,the parties perform full state tomography.
\item Being ensured that the shared state is an entangled state they perform local filtering operations: Alice using operators $\{M_{A}^{(1)},M_{A}^{(2)}\}$ and Bob performing $\{M_{B}^{(1)},M_{B}^{(2)}\}$ where the operators are specified in subsec.\ref{locf}. \\\\
It is known that, in QKD protocols the optical system adopts a prescription wherein the photons are destroyed due to measurements. Hence the intermediate step given below is crucial.\\
\item \textit{Post selection of outcomes:}After completion of measurements, they announce the outcome of measurements in $\{M_A^{(i)}\}$ and $\{M_B^{(j)}\}. $ They post select those qubit pairs for which $M_A^{(1)}$ and $M_B^{(1)}$ clicked. Only these qubit pairs are considered further in the protocol discarding the remaining ones. Each pair of two qubit state selected is denoted by $\rho^{'}.$\\
\item Rest of the protocol runs as usual(subsec.\ref{pros}).\\
\end{itemize}
\subsection{Expression of $Q_{min}^{'}$}
Let us now formulate the expression of $Q_{min}^{'}$, i.e., $Q_{min}$ obtained from post selected state(obtained corresponding to clicking of measurement $M_A^{(1)}$ and $M_B^{(1)}$). Firstly, $P_{succ}$(possibility of measurement $M_A^{(1)}$ and $M_B^{(1)}$ clicking) takes the form:
\begin{equation}\label{keyn2}
    P_{succ}=\textmd{Tr}[M_{A}^{(1)}\otimes M_{B}^{(1)}\rho( M_{A}^{(1)}\otimes M_{B}^{(1)})^{\dagger}],
\end{equation}
where $M_{A}^{(1)},M_{B}^{(1)}$ are given by Eq.(\ref{filter2}). Correlation tensor($T_{\small{filtered}}$, say) of corresponding post selected state($\rho_{\small{filtered}}$, say) obtained from the initial state $\rho^{'}$(Eq.(\ref{st41})) is given by:
\begin{equation}\label{corrte1}
    T_{\small{filtered}}=\left(\begin{array}{ccc}
     \epsilon_1\epsilon_2t_{11}&0&\frac{1}{2}\mathfrak{m}_{1}\epsilon_1(-1+\epsilon_2^2)\\
     &&\\
     0 &\epsilon_1\epsilon_2t_{22}&\frac{1}{2}\mathfrak{m}_{2}\epsilon_1(-1+\epsilon_2^2)\\
     &&\\
  &&\frac{1}{4}(h_{-}-r_{-}-\\
        -\frac{1}{2}\mathfrak{n}_{1}\epsilon_2\,&-\frac{1}{2}\mathfrak{n}_{2}\epsilon_2\,
        &\epsilon_2^2(h_{-}+r_{-})-\\
     +\frac{1}{2}\mathfrak{n}_{1}\epsilon_2\epsilon_1^2&+\frac{1}{2}\mathfrak{n}_{2}\epsilon_2\epsilon_1^2&\epsilon_1^2(h_{+}-r_{+})+\\
     &&\epsilon_2^2(h_{+}+r_{+}))\\
       \end{array} \right)
\end{equation}
where $h_{\pm}$$=$$1\pm\mathfrak{m}_3,$ $r_{\pm}$$=$$\mathfrak{n}_3\pm t_{33}.$ Sum of the singular values of $T_{\small{filtered}}$(Eq.(\ref{corrte1})) being given by trace of the matrix $\sqrt{(T_{\small{filtered}})^{*}T_{\small{filtered}}},$ QBER ($Q$) in the modified protocol is given by:
\begin{equation}\label{keyn}
   Q^{'}_{min}=P_{succ}*\frac{1}{6}(3-2\sqrt{\sum_{i=1}^2(\epsilon_2^2(1-\epsilon_1^2)^2\mathfrak{n}_i^2+4\epsilon_1^2t_{ii}^2)}-\sqrt{\small{\textbf{B}}})
\end{equation}
where $\textbf{B}$ is a function of $\epsilon_1,\epsilon_2$:
\begin{eqnarray}\label{corrte2}
    \small{\textbf{B}}=\epsilon_1^2(1-\epsilon_2^2)^2(\mathfrak{m}_1^2+\mathfrak{m}_2^2)&+&(h_{-}-r_{-}-\epsilon_2^2\nonumber\\
    (h_{-}+r_{-})+\epsilon_1^2(-h_{+}&+&r_{+}+\epsilon_2^2(h_{+}+r_{+})))^2
\end{eqnarray}
Explicit form of success probability $P_{succ}$ is given by:
\begin{equation}\label{psuc}
    P_{\small{succ}}=\frac{1}{h_--r_-+\epsilon_2^2(h_-+r_-)+\epsilon_1^2(h_+-r_++\epsilon_2^2(h_++r_+))}
\end{equation}
 $Q^{'}_{min}$$<$$\mathcal{Q}_0$ can thus be written as:
\begin{equation}\label{corrte3}
2\sqrt{\sum_{i=1}^2(\epsilon_2^2(1-\epsilon_1^2)^2\mathfrak{n}_i^2+4\epsilon_1^2t_{ii}^2)}-\sqrt{\small{\textbf{B}}}>
  3- \mathcal{Q}_0*\frac{6}{P_{succ}}
\end{equation}
Above relation(Eq.(\ref{corrte3})) in turn characterizes the states useful in the QKD after applying suitable local filtering operations of the form given by Eq.(\ref{filter2}).
For further discussion in this section we will refer to this QKD protocol(with filtering operations) as \textit{Modified QKD protocol} while that without any filtering operations as QKD protocol only. There exist two qubit states violating Eq.(\ref{basis9iii}) but satisfying Eq.(\ref{corrte3}) for some values of $\epsilon_1,\epsilon_2$.Next we provide some specific examples in support of our claim.\\
\subsection{Illustrations}
\begin{figure}[htb]
\includegraphics[width=3.3in]{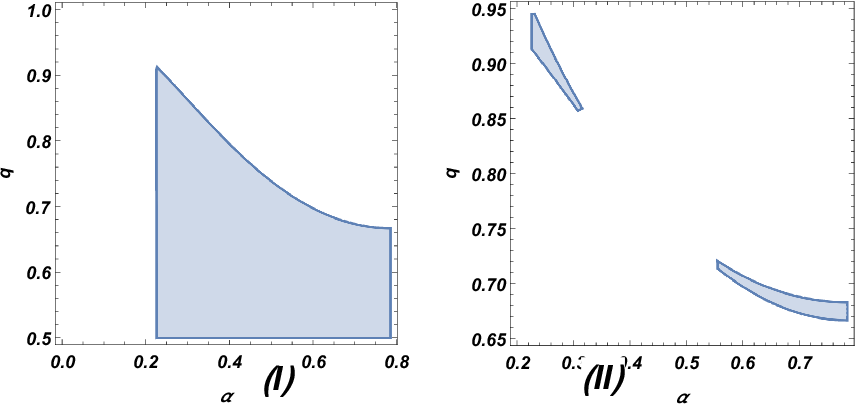}\\
\caption{\emph{Shaded region in subfig.I gives the unsteerable states(Eq.(\ref{s14})) useful in the modified QKD protocol for $\epsilon_1$$=$$.02119$ and $\epsilon_2$$=$$0.02563.$ Similarly, shaded region in subfig.II gives steerable states that are useful in the same modified protocol(i.e., for $\epsilon_1$$=$$.02119$ and $\epsilon_2$$=$$0.02563$). None of these states(both in subfig.I, subfig.II) is useful in the QKD protocol(without filtering operations).}}
\label{fig5}
\end{figure}

\begin{figure}[htb]
\includegraphics[width=3.3in]{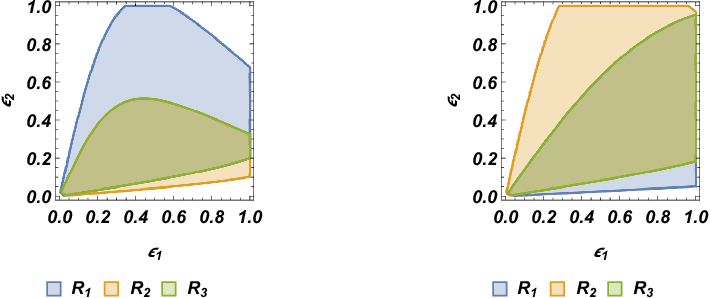}\\
\caption{\emph{In both sub figures parameter space $(\epsilon_1,\epsilon_2)$ characterizing local filtering operations is considered. Shaded regions in each characterize suitable local filtering operations(used in the modified QKD protocol) for some specific states from the class given by Eq.(\ref{s14}). Consider two $F_3$ unsteerable states specified by $(q,\alpha)$$=$$(0.6,0.7)$ and $(q,\alpha)$$=$$(0.5,0.4)$ In sub figure-I, any point in the region $R_1\cup R_2$ gives suitable local filters for the first state whereas that for the second state is given by any point from the region $R_2\cup R_3.$ Next consider two steerable states:$(q,\alpha)$$=$$(0.9,0.26)$ and $(q,\alpha)$$=$$(0.68,0.7)$ In sub figure-II, suitable local filters for these two steerable states are given by any point from region $R_1\cup R_2$ and $R_2\cup R_3$ respectively. }}
\label{fig6}
\end{figure}

 Consider the family of states given by Eq.(\ref{s14}). $F_3$ steerable members from this class which are useless in the QKD violate Eq.(\ref{s142}). Let any such $F_3$ steerable state be used in the modified protocol. For some suitable local filtering operations(specific values of $\epsilon_1,\epsilon_2$ in Eq.(\ref{filter2})), the protocol runs successfully(see fig.\ref{fig5},with the specific values mentioned therein). For a particular instance, consider $\alpha$$=$$0.25,$ $q$$=$$0.9$(Eq.(\ref{s14})) Before filtering $Q_{min}$$=$$0.22839$$>$$\mathcal{Q}_0.$ On using this state in the modified QKD protocol with $\epsilon_1$$=$$0.16119$ and $\epsilon_2$$=$$0.12563,$ $P_{succ}$$\approx$$0.015,$ and by Eq.(\ref{keyn}), $Q_{min}^{'}$$=$$0.13756.$ As $Q_{min}^{'}$$<$$\mathcal{Q}_0,$ so successful key generation takes place in the protocol. For some fixed values of $\alpha,$ range of noise level parameter $q$ for useful states(Eq.(\ref{s141})) are given in Table.(\ref{table:ta1}). Now, for obvious reasons, not all local filtering operations(Eq.(\ref{filter2})) turn out to be useful in the modified protocol. Depending on the state to be used, ($\epsilon_1,\epsilon_2$), parameterizing these operations(Eq.(\ref{filter2})) are to be selected. For the above class of states(Eq.(\ref{s14}) considered, a suitable range of $(\epsilon_1,\epsilon_2)$ is shown in Fig.(\ref{fig6}). \\
\\
Next, let $F_3$ unsteerable states(violating Eq.(\ref{s141})) from this family be used in the modified protocol. Again for some suitable filtering operations made by the users, secure key generation becomes possible for some of these $F_3$ unsteerable states(see Fig.(\ref{fig6})). Some specific instances are given in Table.(\ref{table:ta1}).\\
\begin{center}
\begin{table}[htp]
\caption{Modified QKD protocol with $\epsilon_1$$=$$0.15,$ $\epsilon_2$$=$$0.02563$ is considered. For some specific values of state parameter $\alpha,$ range of other parameter $q$ is specified for which corresponding state is useful for secure key generation in this modified QKD protocol. Second column in the table indicates whether the state used in the protocol violates CJWR inequality or not. Last two instances point out the fact that initially $F_3$ unsteerable states can also be used for the modified QKD protocol.}
\begin{center}
\begin{tabular}{|c|c|c|}
\hline
State&$F_3$&Range of\\
parameter&Steerability&$q$\\
\hline
$\alpha$$=$$0.24$&Steerable&$[0.904,1]$\\
\hline
$\alpha$$=$$0.7$&Steerable&$[0.674,1]$\\
\hline
$\alpha$$=$$0.2$&Unsteerable&$[0.5,1]$\\
\hline
$\alpha$$=$$0.6$&Unsteerable&$[0.52,1]$\\
\hline
\end{tabular}
\end{center}
\label{table:ta1}
\end{table}
\end{center}
\subsection{Other Local Filters}
\begin{figure}[htb]
\includegraphics[width=3.3in]{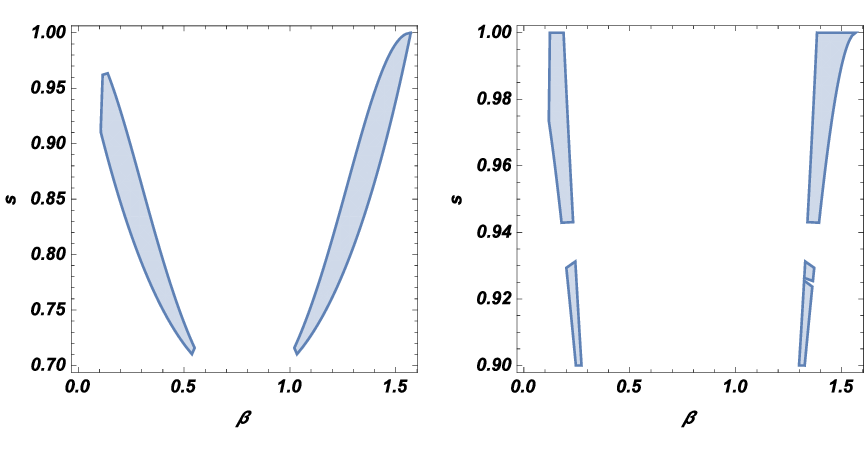}\\
\caption{\emph{Subfig.I gives $F_3$ unsteerable states(Eq.(\ref{gisin1})) useful in the modified QKD protocol. Similarly, shaded region in subfig.II gives steerable states that are useful in the modified protocol. None of these states(in both subfigures) is useful in the QKD protocol(without filtering operations).}}
\label{fig5new}
\end{figure}

Now, as already stated before in sec.\ref{pre}, form of filters Eq.(\ref{filter2})) is not general. Depending on the state provided, other form of filters may also turn out to be suitable in the modified protocol. For instance, consider the well known family of Gisin states \cite{pop2}:
\begin{equation}\label{gisin1}
    \gamma=s |\varphi\rangle\langle\varphi|+\frac{1-s}{2}(|11\rangle\langle11|+|00\rangle\langle00|)
\end{equation}
where $|\varphi\rangle=\cos\beta|01\rangle+\sin\beta|1\rangle,$ with $\beta$$\in$$[0,\frac{\pi}{4}]$ and $0$$\leq $$s$$\leq$$1$. Correlation tensor of this class of states is $\textmd{diag}(s \sin2\beta,s \sin2\beta,1-2s)$. Suitable local filters for this state are of the form \cite{hirsch}:
\begin{eqnarray}\label{filter21}
  F_{A}^{(1)} &=& \sqrt{\tan(\beta)}|0\rangle\langle0|+|1\rangle\langle 1|\\
F_{B}^{(1)} &=& |0\rangle\langle0|+\sqrt{\tan(\beta)}|1\rangle\langle 1|
\end{eqnarray}
Correlation tensor of the post selected state(output corresponding to above filters Eq.(\ref{filter21})) is given by $\frac{1}{1-s+s\sin2\beta}\textmd{diag}(s \sin2\beta,s \sin2\beta,-1+s+s\sin2\beta).$ There exist members from this class of states(Eq.(\ref{gisin1})), which turn out to be useful in our modified QKD protocol(see fig.\ref{fig5new}). For a particular example, consider the state specified by $s$$=$$0.87$ and $\beta$$=$$0.29.$ For this state $Q_{min}$$=$$0.21774.$ Success probability($P_{succ}$) that $F_A^{(1)}\otimes F_B^{(1)}$ will click is $0.18$ and $Q_{min}^{'}$$=$$0.14283.$ The state is thus useful in the modified QKD protocol but cannot be used for secure key generation before filtering.\\
After discussing how our QKD protocol can be modified by allowing the users to apply local filtering operations, we next consider another significant aspect of our protocol.
\section{Absolute Bell-CHSH Local States In Secure Key Generation}\label{abs}
\begin{figure}[htb]
\includegraphics[width=3.3in]{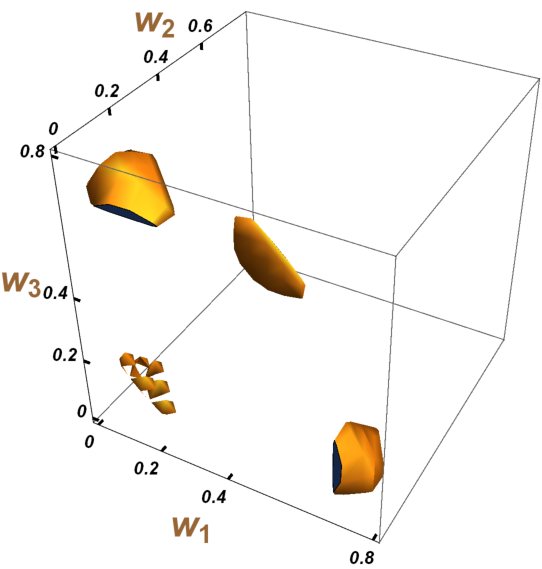}\\
\caption{\emph{Corresponding to any point from the shaded regions of the parameter space $(w_1,w_2,w_3)$, absolutely local Bell diagonal state(Eq.(\ref{belldiag})) is useful in our QKD protocol.}}
\label{fig8}
\end{figure}

Our QKD protocol being semi device independent in nature, entangled state is distributed from some unknown source $\Lambda.$ Let an untrusted third party Eve has access to $\Lambda.$ So if $\rho$ be the state generated from $\Lambda,$ then Eve has access to both the qubits of $\rho.$ Let, Eve measure $\rho$ in some suitable global basis such that $\rho$ transforms into $\rho^{'}$ where $\rho^{'}$ remains entangled but becomes Bell-CHSH local in the new basis. Under control of Eve source $\Lambda$ thus distributes an absolutely Bell-CHSH local state $\rho^{'}$ between Alice and Bob in the protocol. Unlike that of any standard QKD protocol relying on Bell-CHSH violation, our protocol can securely generate key for some of these states. Owing to existence of absolutely Bell-CHSH local $F_3$ steerable two qubit states, our QKD protocol gives advantage over the standard ones. We provide an example below, \\
Consider the family of Bell diagonal states(Eq.(\ref{belldiag})). Parameters of absolutely Bell-CHSH local states from this family satisfy\cite{guoch}:
\begin{eqnarray}\label{belldiag5}
\textmd{Max}_{(i,j,k)}[1-4(w_i-w_i^2-w_i*w_j-w_i*w_k)&&\nonumber\\
-2(w_j+w_k-w_j^2-w_k^2)]&\leq& \frac{1}{2},
\end{eqnarray}
where $(i,j,k)$ denote all possible cyclic permutations(of length three) over $1,2,3:\,(1,2,3),\,(2,3,1)\,(3,1,2).$
Let any such state gets distributed between the two users of the protocol. Some of these states satisfy Eq.(\ref{belldiag2}). Consequently, the protocol runs successfully for them(see Fig.\ref{fig8}).\\
\section{Discussions}\label{conc}
Notion of Bell nonlocality has been rigorously analyzed with respect to the study of QKD in entanglement assisted protocols.
However, whether Bell violation provides a sufficient criterion is a matter of great debate. Present discussion points out the insufficiency of a Bell type inequality in this perspective. CJWR inequality(Eq.(\ref{mon2i})), a Bell type inequality for detecting steerability is used here, to prescribe a QKD protocol. Using notion of $F_3$ steerability, characterization of arbitrary two qubit state is obtained in the context of its usefulness in QKD protocol. Interestingly, any $F_3$ steerable Werner state is useful in QKD protocol. For any amount of violation $\mathcal{V}$ of Eq.(\ref{mon2i}), one can identify whether the state(giving $\mathcal{V}$ amount of violation) is useful in the protocol or not. Such an identification is completely based on the singular values of the correlation tensor of the corresponding state.
\par Furthermore, in case local filtering operations are allowed(before the users perform local bases measurements) in the protocol, some $F_3$ unsteerable states become useful in the modified QKD protocol. The utility of absolutely local Bell-CHSH states in our modified protocol further buttresses our work.
\par Critical quantum bit error rate $\mathcal{Q}_0$(Eq.(\ref{bases7})) in our protocol is greater than that obtained for protocol depending on Bell-CHSH nonlocality($\mathcal{Q}_{0}$$=$$0.14$ \cite{main}). Increased number of measurement settings per party($2$ in \cite{main} and $3$ here) is one of the potent causes for such a contrast.
\par In order to perform entanglement based QKD protocols, one needs to transmit part of quantum systems through noisy channels, which can in turn affect the state of the quantum systems. In this non ideal scenario identification of states offering utility in QKD assumes significance, which our work addresses.
\par In experimental situations, there exist loopholes in testing any correlator based inequality. Owing to experimental imperfections, testing CJWR inequality(Bell-type inequality) may suffer from three major loopholes: locality loopholes(due to hidden communication between the parties\cite{lp1,lp11}), detection loopholes(due to unfair sampling of ensemble which is measured\cite{lp2}) and freedom-of-choice loophole(owing to possible influences from or on selection of measurement settings\cite{lp3}). Our protocol, being based on CJWR violation, these loopholes will exist in any experimental demonstration of the same. Also, classical communication over public channel(for key generation) forms a potent factor of experimental imperfections. Furthermore, in the modified QKD protocol, the parties need to perform full tomography(involving classical communication between the parties) on the quantum state received from the source. After state tomography, in case the state turns out to be an entangled one, local filtering operations are performed followed by usual steps of QKD protocol. The parties need to communicate classically for performing state tomography which may again potentially open up loophole for the protocol. In \cite{lp4}, an experimental demonstration of EPR steering has been provided where each of detection, locality and freedom of choice loophole is closed simultaneously. It will be interesting to explore possible means of closing the loopholes arising in our protocol.
\par As already specified before, the entire analysis in our work is applicable for those QKD protocols where secure key rate is a function of QBER($Q$) only\cite{eck8}. However owing to complexity of practical situations, $\mathfrak{r}_{min}$ may depend on many other factors. So, characterizing arbitrary two qubit states for more general QKDs is a potential source of future research. A generalization of the scheme to include other Bell-like inequalities also warrants attention. Also it will be interesting to analyze the situation when the users of the protocol do not have any knowledge about the dimension of the quantum state distributed by the source.
\section*{Acknowledgement}
Tapaswini Patro would like to acknowledge the support from DST-Inspire fellowship No. DST/INSPIRE Fellowship/2019/IF190357. We are grateful to Shashank Gupta for useful insights.
\section{CONTRIBUTION OF THE AUTHORS}
K. Mukherjee developed main idea of the work, performed the analysis and wrote the paper. N.Ganguly and T.Patro cross checked the findings and also assisted K. Mukherjee in writing the paper. All the authors have read and approved the final manuscript.
\section{Data availability statement}
No Data associated in the manuscript.

\section{Appendix.I}\label{app1}
\underline{\textit{Proof regarding optimization of $ Q $,Eq(\ref{basis6})}}:\\
Let $\vec{\mathfrak{m}}$$ =$$(m_1,m_2,m_3)$ denote an arbitrary direction. Eigen basis operators corresponding to projective measurement along $\vec{\mathfrak{m}}$ are given by: $\{\frac{\mathbb{I}_2+\vec{\mathfrak{m}}.\vec{\sigma}}{2},
\frac{\mathbb{I}_2-\vec{\mathfrak{m}}.\vec{\sigma}}{2}\}.$ Now, as discussed in the main text, in our QKD protocol each of Alice and Bob perform projective measurements along any one of three arbitrary directions: $\vec{u}_1,\vec{u}_2,\vec{u}_3$ for Alice and $\vec{v}_1,\vec{v}_2,\vec{v}_3$ for Bob. So for Alice the collection of measurement basis operators(Eq.(\ref{opes})) is given by:
\begin{equation}\label{ap1}
     \mathcal{O}_{A}^{(j)}=\{\frac{\mathbb{I}_2+\vec{u_j}.\vec{\sigma}}{2},
     \frac{\mathbb{I}_2-\vec{u_j}.\vec{\sigma}}{2}\},\, j=1,2,3
\end{equation}
Similarly, collection of measurement basis operators of Bob is given by:
\begin{equation}\label{ap2}
     \mathcal{O}_{B}^{(j)}=\{\frac{\mathbb{I}_2+\vec{v_j}.\vec{\sigma}}{2},
     \frac{\mathbb{I}_2-\vec{v_j}.\vec{\sigma}}{2}\},\, j=1,2,3
\end{equation}
As discussed in the main text, corresponding to correlated bases of Alice and Bob, the operator bases are given by $\mathcal{O}_{A}^{(j)},\mathcal{O}_{B}^{(j)}(j$$=$$1,2,3).$\\
An arbitrary two qubit state $\rho$(Eq.(\ref{st4})) is shared between Alice and Bob. In case Alice and Bob measure $\vec{u_i}.\vec{\sigma},\vec{v_i}.\vec{\sigma},$ probability of them obtaining mismatching outputs while measuring in correlated bases is given by:
\begin{equation}\label{ap3}
P_j=\frac{1}{4}\sum_{i=0,1}\textmd{Tr}[(\mathbb{I}_2+(-1)^i\vec{u_j}.\vec{\sigma})\otimes
(\mathbb{I}_2+(-1)^{i+1}\vec{v_j}.\vec{\sigma}).\rho],
\end{equation}
$\forall\,j$$=$$1,2,3.$
Using Eq.(\ref{ap3}) the expression for quantum bit error rate $Q$(Eq.(\ref{basis2})) becomes:
\begin{eqnarray}\label{ap4}
    Q&=&\sum_{i=1}^3 P_i\nonumber\\
    &=&\frac{1}{6}(3-\sum_{i=1}^3\sum_{j=1}^3u_{ij}\sum_{k=1}^3w_{jk}v_{ik})\nonumber\\
    &=&\frac{1}{6}(3-\sum_{i=1}^3\vec{u_i}.W\vec{v_i}),
\end{eqnarray}
where $\vec{u_i}$$=$$(u_{i1},u_{i2},u_{i3})^{t}$ and $\vec{v_i}$$=$$(v_{i1},v_{i2},v_{i3}),\forall i$$=$$1,2,3.$
Eq.(\ref{ap4}) gives Eq.(\ref{basis6}).\\

\underline{\textit{Proof of Eq(\ref{basis7})}}:Minimization of $Q$ over all possible bases of the two parties.\\
 \\
Clearly $Q$ is summation of the probability terms appearing in Eq.(\ref{ap3}). Let us introduce a few notations $G_i^{z_1},\,H_i^{z_2},\forall i$$=$$1,2,3$ for ease of use in further calculations: \\
\begin{eqnarray}\label{ap4i}
    G_i^{z_1}&=&\frac{1}{2}(\mathbb{I}_2+(-1)^{z_1}\vec{u_i}.\vec{\sigma}),\,z_1=0,1\\
   H_i^{z_2}&=&\frac{1}{2}(\mathbb{I}_2+(-1)^{z_2}\vec{v_i}.\vec{\sigma}),\,z_2=0,1
\end{eqnarray}
Now, let $L_a$ and $L_b$ denote the local unitary operations such that:
\begin{equation}\label{ap5}
\rho=(L_a\otimes L_b)\rho^{'}(L_a\otimes L_b)^{\dagger}
\end{equation}
Let the unitary operations be specified as follows:
\begin{equation}\label{unit}
    L_{a(b)}=
   \left(\begin{array}{cc}
     l_{a(b)}^{(11)}&l_{a(b)}^{(12)}\\
      l_{a(b)}^{(21)}&l_{a(b)}^{(22)}\\
       \end{array} \right)
\end{equation}
Using Eq.(\ref{ap5}), $\forall j$$=$$1,2,3,$ from Eq.(\ref{ap3}), we get:
\begin{eqnarray}\label{ap6}
    P_j&=&\sum_{i=0,1}\textmd{Tr}[G_j^{i}\otimes
H_j^{i+1}((L_a\otimes L_b)\rho^{'}(L_a\otimes L_b)^{\dagger})]\nonumber\\
&=&\sum_{i=0,1}\textmd{Tr}[(L_a\otimes L_b)^{\dagger}G_j^{i}\otimes
H_j^{i+1}(L_a\otimes L_b)\rho^{'}]\nonumber\\
&=&\sum_{i=0,1}\textmd{Tr}[(L_a^{\dagger}G_j^{i}L_a)
\otimes (L_b^{\dagger}H_j^{i+1}L_b)\,\rho^{'}]
\end{eqnarray}
From Eq.(\ref{ap6}), consider the term $L_a^{\dagger}G_j^{i}L_a$ for $i$$=$$0$(say). Let us now further analyze this term. Using Eqs.(\ref{ap4i},\ref{unit}), we get:
\begin{eqnarray}\label{ap7}
  L_a^{\dagger}G_j^{0}L_a &=&L_a^{\dagger}\frac{1}{2}(\mathbb{I}_2+\vec{u_j}.\vec{\sigma})L_a\nonumber\\
  &=& \frac{1}{2}(\mathbb{I}_2+L_a^{\dagger}\vec{u_j}.\vec{\sigma}L_a) \\
\end{eqnarray}
Now using Eq.(\ref{unit}), $L_a,L_a^{\dagger}$ can be expressed in terms of Pauli matrices($\sigma_1,\sigma_2,\sigma_3$) as follows:
\begin{eqnarray}\label{ap8}
  L_a &=&\frac{1}{2}((l_a^{(11)}+l_a^{(22)})\mathbb{I}_2+(l_a^{(11)}-l_a^{(22)}))
  \sigma_3 \nonumber\\
  +(l_a^{(12)}&+&l_a^{(21)})\sigma_1+\imath(l_a^{(12)}-l_a^{(21)})\sigma_2\nonumber\\
  &&\nonumber\\
  L_a^{\dagger} &=& \frac{1}{2}((\overline{l_a^{(11)}}+\overline{l_a^{(22)}})\mathbb{I}_2+(\overline{l_a^{(11)}}-
 \overline{ l_a^{(22)}})) \sigma_3 \nonumber\\
  +(\overline{l_a^{(12)}}&+&\overline{l_a^{(21)}})\sigma_1+
  \imath(\overline{l_a^{(21)}}-\overline{l_a^{(12)}})\sigma_2
  \end{eqnarray}
with $\overline{l_a^{(ij)}}$ denoting complex conjugate of $l_a^{(ij)}.$
Using Eq.(\ref{ap8}), from Eq.(\ref{ap7}), we get:
\begin{eqnarray}\label{ap9}
 L_a^{\dagger}\vec{u_j}.\vec{\sigma}L_a&=& \left(\begin{array}{cc}
   A_{11}&A_{12}\\
      A_{21}&A_{22}\\
\end{array} \right),\,\textmd{where,}\nonumber \\
 A_{11}&=&
 (l_{a(b)}^{(21)} (u_{j1} - \imath\, u_{j2}) + l_{a(b)}^{(11)} u_{j3}) \overline{
   l_{a(b)}^{(11)}}\nonumber\\
    &+& (l_{a(b)}^{(11)} (u_{j1} + \imath\, u_{j2}) - l_{a(b)}^{(21)} u_{j3}) \overline{l_{a(b)}^{(21)}}\nonumber\\
   A_{12}&=&(l_{a(b)}^{(21)} (u_{j1} - \imath\, u_{j2}) + l_{a(b)}^{(11)} u_{j3}) \overline{
  l_{a(b)}^{(12)}}\nonumber\\
   &+& (l_{a(b)}^{(11)} (u_{j1} + \imath\, u_{j2}) - l_{a(b)}^{(21)} u_{j3}) \overline{l_{a(b)}^{(22)}}\nonumber\\
  A_{21}&=&(l_{a(b)}^{(22)} (u_{j1} - \imath\, u_{j2}) + l_{a(b)}^{(12)} u_{j3}) \overline{
  l_{a(b)}^{(11)}} \nonumber\\
  &+& (l_{a(b)}^{(12)} (u_{j1} + \imath\, u_{j2}) - l_{a(b)}^{(22)} u_{j3}) \overline{l_{a(b)}^{(21)}}\nonumber\\
  A_{22}&=&(l_{a(b)}^{(22)} (u_{j1} - \imath\, u_{j2}) + l_{a(b)}^{(12)} u_{j3}) \overline{
  l_{a(b)}^{(12)}} \nonumber\\
  &+& (l_{a(b)}^{(12)} (u_{j1} + \imath\, u_{j2}) - l_{a(b)}^{(22)} u_{j3}) \overline{l_{a(b)}^{(22)}}
 \end{eqnarray}
$L_a$ being an unitary matrix, $L_a.L_a^{\dagger}$$=$$L_a^{\dagger}.L_a$$=$$\mathbb{I}_2.$ Using that we get:
\begin{eqnarray}\label{ap10}
  |\small{l_a^{(11)}}|^2 +|\small{l_a^{(12)}}|^2 &=& 1 \nonumber\\
  |\small{l_a^{(21)}}|^2 +|\small{l_a^{(22)}}|^2 &=& 1 \nonumber\\
 \small{l_a^{(22)}} \overline{\small{l_a^{(12)}}} &=&-\small{l_a^{(21)}} \overline{\small{l_a^{(11)}}}\nonumber\\
 \small{l_a^{(11)}} \overline{\small{l_a^{(21)}}} &=& - \small{l_a^{(12)}} \overline{\small{l_a^{(22)}}}
\end{eqnarray}
Using above set of relations(Eq.(\ref{ap10})), $L_a^{\dagger}\vec{u_j}.\vec{\sigma}L_a$(Eq.(\ref{ap9})) can be expressed in terms of Pauli matrices as follows:
\begin{equation}\label{ap11}
 L_a^{\dagger}\vec{u_j}.\vec{\sigma}L_a=\vec{u_j}^{'}.\vec{\sigma}
\end{equation}
where $\vec{u_j}^{'}$$=$$(u_{j1}^{'},u_{j2}^{'},u_{j3}^{'}).$ It components are specified as follows:
\begin{eqnarray}\label{ap12}
  u_{j1}^{'} =\frac{1}{2}( \imath \small{u_{j2}} (-\small{l_a^{(22)}} \overline{\small{\small{l_a^{(11)}}}}& -& \small{l_a^{(21)}} \overline{\small{l_a^{(12)}}} +
    \small{l_a^{(12)}} \overline{\small{l_a^{(21)}}} + \small{\small{l_a^{(11)}}} \overline{\small{l_a^{(22)}}})\nonumber\\
     + \small{u_{j1}} (\small{l_a^{(22)}} \overline{\small{\small{l_a^{(11)}}}} &+& \small{l_a^{(21)}} \overline{\small{l_a^{(12)}}} +
    \small{l_a^{(12)}} \overline{\small{l_a^{(21)}}} + \nonumber\\
    \small{\small{l_a^{(11)}}} \overline{\small{l_a^{(22)}}})& +&
 \small{u_{j3}} (-\small{l_a^{(22)}} \overline{\small{l_a^{(21)}}} - \small{l_a^{(21)}} \overline{\small{l_a^{(22)}}}))\nonumber \\
  u_{j2}^{'} =\frac{\imath}{2} ( \small{u_{j1}} (\small{l_a^{(22)}} \overline{\small{\small{l_a^{(11)}}}} &-& \small{l_a^{(21)}} \overline{\small{l_a^{(12)}}} + \small{l_a^{(12)}} \overline{\small{l_a^{(21)}}} - \small{\small{l_a^{(11)}}} \overline{\small{l_a^{(22)}}}) \nonumber\\
    + \small{u_{j2}} (\small{l_a^{(22)}} \overline{\small{\small{l_a^{(11)}}}} &-& \small{l_a^{(21)}} \overline{\small{l_a^{(12)}}} -
    \small{l_a^{(12)}} \overline{\small{l_a^{(21)}}} +\nonumber\\
     \small{\small{l_a^{(11)}}} \overline{\small{l_a^{(22)}}}) &+&
 \imath \small{u_{j3}} (-\small{l_a^{(22)}} \overline{\small{l_a^{(21)}}} + \small{l_a^{(21)}} \overline{\small{l_a^{(22)}}}))\nonumber \\
   u_{j3}^{'} = \small{u_{j1}} (-\small{l_a^{(22)}} \overline{\small{l_a^{(12)}}}& -& \small{l_a^{(12)}} \overline{\small{l_a^{(22)}}}) +
 \imath \small{u_{j2}}
  (\small{l_a^{(22)}} \overline{\small{l_a^{(12)}}}\nonumber\\
   - \small{l_a^{(12)}} \overline{\small{l_a^{(22)}}})& +&
 \small{u_{j3}} (-1 + 2 \small{l_a^{(22)}} \overline{\small{l_a^{(22)}}})
\end{eqnarray}
On simplification of the above equations, each component of $\vec{u_j}^{'}$ turns out to be real quantity:
\begin{eqnarray}\label{ap13}
   u_{j1}^{'} &=&\small{u_{j2}} (\textmd{\small{Im}}[\small{l_a^{(22)}} \overline{\small{\small{l_a^{(11)}}}}]+ \textmd{\small{Im}}[\small{l_a^{(21)}} \overline{\small{l_a^{(12)}}}]) +\nonumber\\
 \small{u_{j1}} (\textmd{\small{Re}}[\small{l_a^{(22)}} \overline{\small{\small{l_a^{(11)}}}}]& +& \textmd{\small{Re}}[\small{l_a^{(21)}} \overline{\small{l_a^{(12)}}}]) -
 2 \small{u_{j3}} \textmd{\small{Re}}[\small{l_a^{(22)}} \overline{\small{l_a^{(21)}}}] \nonumber\\
 &&\nonumber\\
   u_{j2}^{'} &=&\small{u_{j1}} (-\textmd{\small{Im}}[\small{l_a^{(22)}} \overline{\small{\small{l_a^{(11)}}}}] + \textmd{\small{Im}}[\small{l_a^{(21)}} \overline{\small{l_a^{(12)}}}]) +\nonumber\\
 2 \small{u_{j3}} \textmd{\small{Im}}[\small{l_a^{(22)}} \overline{\small{l_a^{(21)}}}] &+&
 \small{u_{j2}} (\textmd{\small{Re}}[\small{l_a^{(22)}} \overline{\small{\small{l_a^{(11)}}}}] - \textmd{\small{Re}}[\small{l_a^{(21)}} \overline{\small{l_a^{(12)}}}]) \nonumber \\
 &&\nonumber\\
   u_{j3}^{'} = -1 + 2 |\small{l_a^{(22)}}|^2 &-& 2 \small{u_{j2}} \textmd{\small{Im}}[\small{l_a^{(22)}} \overline{\small{l_a^{(12)}}}] -
 2 \small{u_{j1}} \textmd{\small{Re}}[\small{l_a^{(22)}} \overline{\small{l_a^{(12)}}}]
\end{eqnarray}
Using above relations, length of $\vec{u_j}^{'}$ turns out to be $1.$\\
So, in totality, Eq.(\ref{ap7}) can be expressed as:
\begin{equation}\label{ap13}
 L_a^{\dagger}G_j^{0}L_a = \frac{1}{2}(\mathbb{I}_2+\vec{u_j}^{'}.\vec{\sigma})
\end{equation}
where $\vec{u_j}^{'}$ is a unit length real vector for each of $j$$=$$1,2,3.$\\
Analogous argument can be put for each of $L_a^{\dagger}G_j^{1}L_a,\, L_a^{\dagger}H_j^{0}L_a$ and $L_a^{\dagger}H_j^{1}L_a.$ So $P_j$(Eq.(\ref{ap6})) now becomes:
\begin{equation}\label{ap14}
\small{P_j}=\frac{1}{4}\sum_{i=0,1}\textmd{\small{Tr}}\small{[(\mathbb{I}_2+(-1)^i\vec{u_j}^{'}.\vec{\sigma})\otimes
(\mathbb{I}_2+(-1)^{i+1}\vec{v_j}^{'}.\vec{\sigma}).\rho^{'}]},
\end{equation}
where $\vec{v_j}^{'}.\vec{\sigma}$$=$$L_b^{\dagger}\vec{v_j}.\vec{\sigma}L_b.$
Using Eq.(\ref{ap14}), $Q$ now becomes:
\begin{eqnarray}\label{ap15}
  Q&=&\frac{1}{6}(3-\sum_{i=1}^3\sum_{j=1}^3u_{ij}^{'}\sum_{k=1}^3T_{jk}v_{ik}^{'})\nonumber\\
    &=&\frac{1}{6}(3-\sum_{i=1}^3\vec{u_i}^{'}.T\vec{v_i}^{'})\nonumber\\
    &=&\frac{1}{6}(3-\sum_{i=1}^3\langle\vec{u_i}^{'},T\vec{v_i}^{'}\rangle)
\end{eqnarray}
where $T$$=\textmd{diag}(t_{11},t_{22},t_{33})$ is the correlation tensor of $\rho^{'}$(Eq.(\ref{st41}))
To prove Eq.(\ref{basis7}), we now need to minimize $Q$(Eq.\ref{ap15}) over all possible measurement directions $\vec{u_j}^{'},\vec{v_j}^{'}(j$$=$$1,2,3).$ \\

\begin{eqnarray}\label{ap15y}
|\langle\vec{u_i}^{'},T\vec{v_i}^{'}\rangle| &\leq& ||\vec{u_i}^{'}|| ||T\vec{v_i}^{'}||\,\forall i=1,2,3 \nonumber\\
 &=& ||T\vec{v_i}^{'}||\,\,\textmd{as}\, ||\vec{u}_i^{'}||=1\nonumber\\
 -||T\vec{v_i}^{'}||&\leq&\langle\vec{u_i}^{'},T\vec{v_i}^{'}\rangle \leq ||T\vec{v_i}^{'}||\nonumber\\
 \textmd{So}, \sum_{i=1}^3\langle\vec{u_i}^{'},T\vec{v_i}^{'}\rangle&\leq &\sum_{i=1}^3 ||T\vec{v_i}^{'}||
\end{eqnarray}
Hence, by Eqs.(\ref{ap15},\ref{ap15y}), we get:
\begin{equation}\label{ap15z}
    Q\geq \frac{1}{6}(3-\sum_{i=1}^3||T\vec{v_i}^{'}||)
\end{equation}
As said in the main text, Alice is not allowed to perform measurements in mutually unbiased basis whereas Bob performs measurement in mutually unbiased bases(MUBs). Now for local dimension $d$$=$$2,$ up to global phase factor, there exist three possible MUBs\cite{mub1}:$\{|0\rangle,|1\rangle\},\,\{\frac{|0\rangle\pm|1\rangle}{2}\}$ and $\{\frac{|0\rangle\pm \imath\,|1\rangle}{2}\}.$ Collection of possible operator bases for each of Alice and Bob are enlisted in Table.\ref{table:ta3}. Minimization of $Q$ is now performed over all these measurement operators.\\
\begin{center}
\begin{table}[htp]
\caption{ All possible mutually unbiased operator bases for local dimension $2$ are specified here for Bob. Corresponding direction $\vec{m}$ of projective measurement $\vec{m}.\vec{\sigma}$ is given. Clearly, up to global phase, the three possible MUB operator bases are given corresponding to directions $\vec{m_1},\vec{m_3}$ and $\vec{m_5}.$ Each of three measurement directions of Bob is thus given by third column of the table, i.e., $\forall j$$=$$1,2,3,\,\vec{v_j}^{'}$$\in$$\{\vec{m_1},\vec{m_2},\vec{m_3},\vec{m_4},
\vec{m_5},\vec{m_6}\}.$ As Bob performs three measurement in MUBs, so if $\vec{v_1}^{'}$ is one of $\vec{m_1}$ or $\vec{m_2}$(say), then $\vec{v_2}^{'}$$\neq$$\vec{m_1},\vec{m_2}.$ It can be any one of from ($\vec{m_3}$,$\vec{m_4}$) or from ($\vec{m_5}$,$\vec{m_6}$). If say $\vec{v_2}^{'}$$=\vec{m_3},$ then $\vec{v_3}^{'}$ is any one of $\vec{m_5}$ or $\vec{m_6}.$ }
\begin{center}
\begin{tabular}{|c|c|c|}
\hline
$i$&$\mathcal{O}_i$&$\vec{m_i}$ \\
\hline
$1$&$\{|0\rangle\langle 0|,|1\rangle\langle 1|\}$&$\vec{m_1}$$=$$(0,0,1)$\\
&&\\
\hline
$2$&$\{-|0\rangle\langle 0|,-|1\rangle\langle 1|\}$&$\vec{m_2}$$=$$(0,0,-1)$\\
&&\\
\hline
$3$&$\{\frac{1}{2}(|0\rangle\langle 0|+|0\rangle\langle 1|+|1\rangle\langle 0|+|1\rangle\langle 1|),$&$\vec{m_3}$$=$$(1,0,0)$\\
&$\frac{1}{2}(|0\rangle\langle 0|-|0\rangle\langle 1|-|1\rangle\langle 0|+|1\rangle\langle 1|)\}$&\\
&&\\
\hline
$4$&$\{-\frac{1}{2}(|0\rangle\langle 0|+|0\rangle\langle 1|+|1\rangle\langle 0|+|1\rangle\langle 1|),$&$\vec{m_4}$$=$$(-1,0,0)$\\
&$-\frac{1}{2}(|0\rangle\langle 0|-|0\rangle\langle 1|-|1\rangle\langle 0|+|1\rangle\langle 1|)\}$&\\
&&\\
\hline
$5$&$\{\frac{1}{2}(|0\rangle\langle 0|+\imath|0\rangle\langle 1|+\imath|1\rangle\langle 0|-|1\rangle\langle 1|),$&$\vec{m_5}$$=$$(0,1,0)$\\
&$\frac{1}{2}(|0\rangle\langle 0|-\imath|0\rangle\langle 1|-\imath|1\rangle\langle 0|-|1\rangle\langle 1|)\}$&\,\\
&&\\
\hline
$6$&$\{-\frac{1}{2}(|0\rangle\langle 0|+\imath|0\rangle\langle 1|+\imath|1\rangle\langle 0|-|1\rangle\langle 1|),$&$\vec{m_6}$$=$$(0,-1,0)$\\
&$-\frac{1}{2}(|0\rangle\langle 0|-\imath|0\rangle\langle 1|-\imath|1\rangle\langle 0|-|1\rangle\langle 1|)\}$&\,\\
&&\\
\hline
\end{tabular}
\end{center}
\label{table:ta3}
\end{table}
\end{center}
Now, all $t_{ii}$$\geq$$0.$ So minimum value of R.H.S. of Eq.(\ref{ap15z}) is obtained for $\vec{v^{'}_1}$$=$$\vec{m_3},$ $\vec{v^{'}_2}$$=$$\vec{m_5}$ and $\vec{v^{'}_3}$$=$$\vec{m_1}$(see Table\ref{table:ta3}):
\begin{equation}\label{ap16}
     Q=\frac{1}{6}(3-t_{11}-t_{22}-t_{33}).
\end{equation}
Expression for $Q_{min}$(Eq.(\ref{basis7})) is thus obtained.\\

\section{Appendix.II}\label{app2}
\textit{\underline{Proof regarding critical error rate, Eq.(\ref{bases7})}:}\\
It may be noted that minimizing $Q_{min}$(Eq.(\ref{basis7})), is equivalent to maximizing the following expression:
\begin{equation}\label{bp1i}
    f(t_{11},t_{22},t_{33})=t_{11}+t_{22}+t_{33}.
\end{equation}
where $f(t_{11},t_{22},t_{33})$ is a symmetric function of eigen values of the correlation tensor of $\rho$(Eq.\ref{st4}). Here we use Lagrange multiplier's method to maximize $f(t_{11},t_{22},t_{33})$ subjected to the constraint provided by Eq.(\ref{basis7i}).\\
Let $\gamma_1$ be the lagrange multiplier. Consider the following function:
\begin{equation}\label{bp2}
     F_1(t_{11},t_{22},t_{33},\gamma_1)=t_{11}+t_{22}+t_{33}+\gamma_1( t_{11}^{2}+ t_{22}^{2}+t_{33}^2-1).
\end{equation}
Partial differentiation of $ F_1(t_{11},t_{22},t_{33},\gamma_1)$ with respect to each of the variables $t_{11},t_{22},t_{33}$ gives
\begin{equation}\label{bp3}
    \frac{\partial F_1}{\partial t_{ii}}=1+2\gamma_1t_{ii},\,i=1,2,3.
\end{equation}
Critical point is then given by $\frac{\partial F_1}{\partial t_{ii}}$$=$$0$ which in turn gives:
\begin{equation}\label{bp4}
    t_{ii}=-\frac{1}{2\gamma_1},\,i=1,2,3.
\end{equation}
Using Eq.(\ref{bp4}), in Eq.(\ref{basis7i}), we get:
\begin{equation}\label{bp5}
    \gamma_1=\pm\frac{\sqrt{3}}{2}
\end{equation}
Now, for this case, as all $t_{ii}$$\geq$$0,$ so by Eqs.(\ref{bp3},\ref{bp4}), $\gamma_1$$=$$-\frac{\sqrt{3}}{2}.$ Critical point($K_1$,say) is thus given by
\begin{equation}\label{bp6}
K_1=(\frac{1}{\sqrt{3}},\frac{1}{\sqrt{3}},\frac{1}{\sqrt{3}})
\end{equation}
Now calculating second order differential of $ F_1(t_{11},t_{22},t_{33},\gamma_1),$ we get:
\begin{eqnarray}\label{bp7}
    d^2 F_1(t_{11},t_{22},t_{33},\gamma_1)&=&\sum_{i,j=1}^3\frac{\partial^2F_1}{\partial t_{ii}\partial t_{jj}}(d t_{ii}d t_{jj})\nonumber\\
    &=&2\gamma_1(d t_{ii})^2\nonumber\\
    &=&-\frac{1}{\sqrt{3}}\nonumber\\
    &<&0
\end{eqnarray}
Eq.(\ref{bp6}) points out that $d^2F $ turns out to be negative at all points. Hence, $K_1$ is the maxima of the objective function $f$(Eq.(\ref{bp1i})), maximum value being given by:
\begin{equation}\label{bp8}
    f(\frac{1}{\sqrt{3}},\frac{1}{\sqrt{3}},\frac{1}{\sqrt{3}})=\sqrt{3}
\end{equation}
Eqs.(\ref{basis7},\ref{bp8}) in turn gives Eq.(\ref{bases7}).
\end{document}